\documentclass[english]{book}

\usepackage[utf8]{inputenc}
\usepackage{graphicx}
\usepackage[sectionbib]{natbib}
\usepackage{chapterbib}
\usepackage{wileyvch}
\crop[noinfo]
\author{John Q.Public}\title{The Public Book}

\begin{document}
\frontmatter
\mainmatter

\chapterauthor{Richard Spinney and Ian Ford} \chapter{Fluctuation relations: a pedagogical overview}

Department of Physics and Astronomy and London Centre for Nanotechnology, \break
University College London, Gower Street, London WC1E 6BT, U.K.\footnote{chapter contributed to R.Klages, W.Just, C.Jarzynski (Eds.), Nonequilibrium Statistical Physics of Small Systems: Fluctuation Relations and Beyond (Wiley-VCH, Weinheim, 2012; ISBN 978-3-527-41094-1)}

\section{Preliminaries}

Ours is a harsh and unforgiving universe, and not just in the little
matters that conspire against us. Its complicated rules of evolution
seem unfairly biased against those who seek to predict the future.
Of course, if the rules were simple, then there might be no universe
of any complexity worth considering. Perhaps richness of behaviour
only emerges because each component of the universe interacts with
many others, and in ways that are very sensitive to details: this
is the harsh and unforgiving nature. In order to predict the future,
we have to take into account all the connections between the components,
since they might be crucial to the evolution, and furthermore, we
need to know everything about the present in order to predict the
future: both of these requirements are in most cases impossible. Estimates
and guesses are not enough: unforgiving sensitivity to the detail
very soon leads to loss of predictability. We see this in the workings
of a weather system. The approximations that meteorological services
make in order to fill gaps in understanding, or initial data, eventually
make the forecasts inaccurate.

So a description of the dynamics of a complex system is likely to
be incomplete and we have to accept that predictions will be uncertain.
If we are careful in the modelling of the system, the uncertainty will
grow only slowly. If we are sloppy in our model building or initial
data collection, it will grow quickly. We might expect the predictions
of any incomplete model to tend towards a state of general ignorance,
whereby we cannot be sure about anything: rain, snow, heatwave or
hurricane. We must expect there to be a spread, or fluctuations, in
the outcomes of such a model.

This discussion of the growth of uncertainty in predictions has a
bearing on another matter: the apparent \emph{irreversibility} of
all but the most simple physical processes. This refers to our inability
to drive a system exactly backwards by reversing the external forces
that guide its evolution. Consider the mechanical work required to
compress a gas by a piston in a cylinder. We might hope to see the expended
energy returned when we stop pushing and allow the gas to drive the piston all
the way back to the starting point: but not all will be returned.
The system seems to mislay some energy to the benefit of the wider
environment. This is the familiar process of friction. The one-way
dissipation of energy during mechanical processing is an example of
the famous second law of thermodynamics. But the process is actually
rather mysterious: what about the underlying reversibility of Newton's
equations of motion? Why is the leakage of energy one-way?

We might suspect that a failure to engineer the exact reversal of
a compression is simply a consequence of a lack of control over all
components of the gas and its environment: the difficulty in setting
things up properly for the return leg implies the virtual impossibility
of retracing the behaviour. So we might not expect to be able to retrace
exactly. But why do we not sometimes see `antifriction'? A clue might
be seen in the relative size and complexity of the system
and its environment. The smaller system is likely to evolve in a more complicated fashion as a result of the coupling, whilst we might expect the larger environment to be much less affected. There is a disparity in the effect of the coupling on each participant, and it is believed that this is responsible for the apparent one-way nature of friction. It is possible to implement these ideas by modelling the behaviour of a system using uncertain,
or stochastic dynamics. The probability of observing a reversal of
the behaviour on the return leg can be calculated explicitly and it turns
out that the difference between probabilities of observing a particular compression and seeing its reverse on the return leg leads to a measure of the irreversibility
of natural processes. The second law is then a rather simple
consequence of the dynamics. A similar asymmetric treatment of the effect on a system of coupling to a large environment is possible using deterministic and reversible non-linear dynamics. In both cases, Loschmidt's paradox, the apparent breakage of time reversal symmetry for thermally constrained systems, is evaded, though for different reasons.

This chapter describes the so-called \emph{fluctuation relations},
or \emph{theorems}  \cite{Evans93,Evans94,Gallavotti95,Jarzynski97,Crooks99},
that emerge from the analysis of a physical system interacting with
its environment, and which provide the structure that leads to the
conclusion just outlined. They can quantify unexpected outcomes in
terms of the expected. They apply on microscopic as well as macroscopic
scales, and indeed their consequences are most apparent when applied
to small systems. They can be derived on the basis of a rather natural measure of irreversibility, just alluded to,
that offers an interpretation of the second law and the associated
concept of entropy production. The dynamical rules that control the
universe might seem harsh and unforgiving, but they can also be charitable,
and from them have emerged fluctuation relations that seem to provide
a better understanding of entropy, uncertainty, and the limits of
predictability.

This chapter is structured as follows. In order to provide a context
for the fluctuation relations suitable for newcomers to the field we begin with a brief summary of
thermodynamic irreversibility, and then describe how stochastic dynamics
might be modelled. We use a framework based on stochastic, rather than deterministic dynamics since developing both themes here might not provide the most succinct pedagogical introduction. Nevertheless, we refer to the deterministic framework briefly later on, to emphasise its equivalence. We discuss the identification of entropy production
with the degree of departure from dynamical reversibility, and then
take a careful look at the developments that follow, which include the
various fluctuation relations, and consider how the second law might
not operate as we expect. We illustrate the fluctuation relations
using simple analytical models, as an aid to understanding. We conclude
with some final remarks, but the broader implications are to be found
elsewhere in this book, for which we hope this chapter will act as
a helpful background.

\section{Entropy and the second law\label{sec:Entropy-and-the}}

Ignorance and uncertainty has never been an unusual state of affairs
in human perception. In mechanics, Newton's laws of motion provided
tools that seemed to dispel some of the haze: here were mathematical
models that enabled the future to be foretold! They inspired attempts to predict future behaviour in other fields, particularly in
thermodynamics, the study of systems through which matter and energy
can flow. The particular focus in the early days of the field was
the heat engine, a device whereby fuel, and the heat it can generate,
can be converted into mechanical work. Its operation was discovered
to produce a quantity called entropy, that could characterise the
efficiency with which energy in the fuel could be converted into motion.
Indeed entropy seemed to be generated whenever heat or matter flowed.
The second law of thermodynamics famously states that the total entropy
of the evolving universe is always increasing. But this statement
still attracts discussion, more than 150 years after its introduction.
We do not debate the meaning of Newton's second law any more, so why
is the second law of thermodynamics so controversial?

Well, it is hard to understand how there can be a physical quantity
that never decreases. Such a statement demands the breakage of the
principle of time reversal symmetry, a difficulty referred to as Loschmidt's paradox. Newton's equations of motion
do not specify a preferred direction in which time evolves. Time is
a coordinate in a description of the universe and it is just convention
that real world events take place while this coordinate increases.
Given that we cannot actually run time backwards, we can demonstrate
this symmetry in the following way. A sequence of events that takes
place according to time reversal symmetric equations can be inverted
by instantaneously reversing all the velocities of all the participating
components and then proceeding forward in time once again, suitably
reversing any external protocol of driving forces, if necessary. The
point is that any evolution can be imagined in reverse, according
to Newton. We therefore don't expect to observe any quantity that
only ever increases with time. This is the essence of Loschmidt's
objection to Boltzmann's  \cite{Cercignani98} mechanical interpretation
of the second law.

But nobody has been able to initiate a heat engine such that it sucks
exhaust gases back into its furnace and combines them into fuel. The
denial of such a spectacle is empirical evidence for the operation
of the second law, but it is also an expression of Loschmidt's paradox.
Time reversal symmetry is broken by the apparent illegality of entropy-consuming
processes, and that seems unacceptable. Perhaps we should not blindly
accept the second law in the sense that has traditionally been ascribed
to it. Or perhaps there is something deeper going on. Furthermore,
a law that only specifies the sign of a rate of change sounds rather
incomplete.

But what has emerged in the last two decades or so is the realisation
that Newton's laws of motion, when supplemented by the acceptance
of uncertainty in the way systems behave, brought about by roughly
specified interactions with the environment, can lead quite naturally
to a quantity that grows with time, namely uncertainty itself. It
is reasonable to presume that incomplete models of the evolution of
a physical system will generate additional uncertainty in the reliability of the description
of the system as they are evolved. If the velocities were all instantaneously
reversed, in the hope that a previous sequence of events might be
reversed, uncertainty would continue to grow within such a model.
We shall need to quantify this vague notion of uncertainty, of course.
Newton's laws on their own are time reversal symmetric, but intuition
suggests that the injection and evolution of configurational uncertainty
would break the symmetry. Entropy production might therefore be equivalent
to the leakage of our confidence in the predictions of an incomplete
model: an interpretation that ties in with prevalent ideas of entropy
as a measure of information.

Before we proceed further, we need to remind ourselves about the phenomenology
of irreversible classical thermodynamic processes \cite{Kondepudi08}.
A system possesses energy $E$ and can receive additional incremental
contributions in the form of heat $dQ$ from a heat bath at temperature
$T$, and work $dW$ from an external mechanical device that might
drag, squeeze or stretch the system. It helps perhaps to view $dQ$
and $dW$ roughly as increments in kinetic and in potential energy,
respectively. We write the first law of thermodynamics (energy conservation)
in the form $dE=dQ+dW$. The second law is then traditionally given
as Clausius' inequality:
\begin{equation}
\oint\frac{dQ}{T}\le0,\label{eq:2}
\end{equation}
where the integration symbol means that the system is taken around
a cycle of heat and work transfers, starting and ending in thermal
equilibrium with the same macroscopic system parameters, such as temperature
and volume. The temperature of the heat bath might change
with time, though by definition and in recognition of its presumed large size it always remains in thermal equilibrium,
and so might the volume and shape imposed upon the system during the
process. We can also write the second law for an incremental thermodynamic
process as:
\begin{equation}
dS_{\mathrm{tot}}=dS+dS_{\mathrm{med}},\label{eq:4}
\end{equation}
 where each term is an incremental entropy change, the system again
starting and ending in equilibrium. The change in system entropy is
denoted $dS$ and the change in entropy of the heat bath, or surrounding
medium, is defined as
\begin{equation}
dS_{\mathrm{med}}=-\frac{dQ}{T},\label{eq:3}
\end{equation}
 such that $dS_{\mathrm{tot}}$ is the total entropy change of the
two combined (the `universe'). We see that equation (\ref{eq:2})
corresponds to the condition $\oint dS_{\mathrm{tot}}\ge0$, since
$\oint dS=0$. A more powerful reading of the second law is that
\begin{equation}
dS_{\mathrm{tot}}\ge0,\label{eq:5}
\end{equation}
 for any incremental segment of a thermodynamic process, as long as it starts and ends in equilibrium. An equivalent
expression of the law would be to combine these statements to write
$dW-dE+TdS\ge0$, from which we conclude that the \emph{dissipative}
work (sometimes called irreversible work) in an isothermal process
\begin{equation}
dW_{d}=dW-dF\label{eq:5b}
\end{equation}
 is always positive, where $dF$ is a change in Helmholtz free energy.
We also might write $dS=dS_{\mathrm{tot}}-dS_{\mathrm{med}}$ and
regard $dS_{\mathrm{tot}}$ as a contribution to the change in entropy
of a system that is not associated with a flow of entropy from the
heat bath, the $dQ/T$ term. For a thermally isolated system, where
$dQ=0$, we have $dS=dS_{\mathrm{tot}}$ and the second law then says
that the system entropy increase is due to `internal' generation;
hence $dS_{\mathrm{tot}}$ is sometimes  \cite{Kondepudi08} denoted
$dS_{i}$.

Boltzmann tried to explain what this ever-increasing quantity might
represent at a microscopic level \cite{Cercignani98}. He considered
a thermally isolated gas of particles interacting through pairwise
collisions within a framework of classical mechanics. The quantity
\begin{equation}
H(t)=\int f(\mathbf{v},t)\ln f(\mathbf{v},t)d\mathbf{v},\label{eq:5a}
\end{equation}
 where $f(\mathbf{v},t)d\mathbf{v}$ is the population of particles
with a velocity in the range $d\mathbf{v}$ about $\mathbf{v}$, can
be shown to decrease with time, or remain constant if the population
is in a Maxwell-Boltzmann distribution characteristic of thermal equilibrium.
Boltzmann obtained this result by assuming that the collision rate
between particles at velocities $\mathbf{v}_{1}$ and $\mathbf{v}_{2}$
is proportional to the product of populations at those velocities,
namely $f(\mathbf{v}_{1},t)f(\mathbf{v}_{2},t)$. He proposed that
$H$ was proportional to the negative of system entropy and that his
so-called $H$-theorem provides a sound microscopic and mechanical
justification for the second law. Unfortunately, this does not hold
up. As Loschmidt pointed out, Newton's laws of motion cannot lead
to a quantity that always decreases with time: $dH/dt\le0$ would
be incompatible with the principle of time reversal symmetry that
underlies the dynamics. The $H$-theorem does have a meaning, but
it is statistical: the decrease in $H$ is an expected, but not guaranteed
result. Alternatively, it is a correct result for a dynamical system that does not adhere to time reversal symmetric equations of motion. The neglect of correlation between the velocities of colliding
particles, both in the past and in the future, is where the model
departs from Newtonian dynamics.

The same difficulty emerges in another form when, following Gibbs,
it is proposed that the entropy of a system might be viewed as a property
of an ensemble of many systems, each sampled from a probability density
$P(\left\{ \mathbf{x},\mathbf{v}\right\} )$, where $\left\{ \mathbf{x},\mathbf{v}\right\} $
denotes the positions and velocities of all the particles in a system.
Gibbs wrote
\begin{equation}
S_{{\rm Gibbs}}=-k_{B}\int P\left(\left\{ \mathbf{x},\mathbf{v}\right\} \right)\ln P\left(\left\{ \mathbf{x},\mathbf{v}\right\} \right)\prod d\mathbf{x}d\mathbf{v},\label{eq:5b-1}
\end{equation}
 where $k_{B}$ is Boltzmann's constant and the integration is over
all phase space. The Gibbs representation of entropy is compatible
with all of classical equilibrium thermodynamics. But the probability
density $P$ for an isolated system should evolve in time according
to Liouville's theorem, in such a way that $S_{{\rm Gibbs}}$
is a \emph{constant} of the motion. How, then, can the entropy of
an isolated system, such as the universe, increase? Either equation
(\ref{eq:5b-1}) is only valid for equilibrium situations, something
has been left out, or too much has been assumed.

The resolution of this problem is that Gibbs' expression can represent
thermodynamic entropy, but only if $P$ is not taken to provide an exact
representation of the state of the universe, or if you wish, of an
ensemble of universes. At the very least, practicality requires us
to separate the universe into a system about which we might know and
care a great deal, and an environment with which the system interacts that is much less precisely monitored.
This indeed is one of the central principles of thermodynamics.
We are obliged by this incompleteness to represent the probability of environmental details
in a so-called coarse-grained fashion, which has the effect that the
probability density appearing in Gibbs' representation of the \emph{system}
entropy evolves not according to Liouville's equations, but to versions
with additional terms that represent the effect of an uncertain environment
upon an open system. This then allows $S_{{\rm Gibbs}}$ to change,
the detailed nature of which will depend on exactly how the environmental forces are represented.

For an isolated system however, an increase in $S_{{\rm Gibbs}}$ will emerge
only if we are obliged to coarse-grain some aspect of the system itself.
This line of development could be considered rather unsatisfactory, since it makes the entropy of an isolated system grain-size dependent, and alternatives may be imagined where the entropy of an isolated system is represented by something other than $S_{{\rm Gibbs}}$. The reader is directed to the literature \cite{Evans2011} for further consideration of this matter. However, in this chapter, we shall concern ourselves largely with entropy generation brought about by systems in contact with coarse-grained environments described using stochastic forces, and within such a framework the Gibbs' representation of system entropy will suffice.

We shall discuss a stochastic representation of the additional terms in the system's dynamical equations
in the next section, but it is important to note that a deterministic
description of environmental effects is also possible, and it might perhaps
be thought more natural. On the other hand, the development using
stochastic environmental forces is in some ways easier to present.
But it should be appreciated that some of the early work on fluctuation
relations was developed using deterministic so-called thermostats \cite{Evans93,Evans02},
and that this theme is represented briefly in section \ref{determ}, and elsewhere in this book.

\section{Stochastic dynamics}

\subsection{Master equations}

We pursue the assertion that sense can be made of the second law, its realm of applicability and its failings, when
Newton's laws are supplemented by the explicit inclusion of a developing
configurational uncertainty. The deterministic rules of evolution
of a system need to be replaced by rules for the evolution of the
\emph{probability} that the property should take a particular configuration.
We must first discuss what we mean by probability. Traditionally it
is the limiting frequency that an event might occur amongst a large
number of trials. But there is also a view that probability represents
a distillation, in numerical form, of the best judgement or belief
about the state of a system: our information \cite{Jaynes03}. It is
a tool for the evaluation of \emph{expectation} values of system properties,
representing what we expect to observe based on information about
a system. Fortunately, the two interpretations lead to laws for the
evolution of probability that are of similar form.

So let us derive equations that describe the evolution of probability
for a simple case. Consider a random walk in one dimension, where
a step of variable size is taken at regular time intervals \cite{Risken89,Gardiner09,vanKampen07}.
We write the \emph{master equation} describing such a \emph{stochastic
process}:
\begin{equation}
\mathcal{P}_{n+1}(x_{m})=\sum_{m^{\prime}=-\infty}^{\infty}T_{n}(x_{m}-x_{m^{\prime}}\vert x_{m^{\prime}})\mathcal{P}_{n}(x_{m^{\prime}}),\label{eq:6}
\end{equation}
 where $\mathcal{P}_{n}(x_{m})$ is the probability that the walker is at position
$x_{m}$ at timestep $n$, and $T_{n}(\Delta x\vert x)$ is the transition
probability for making a step of size $\Delta x$ in timestep $n$
given a starting position of $x$. The transition probability may be considered to represent the effect of the environment on the walker.
We presume that Newtonian forces cause the move to be made, but we
do not know enough about the environment to model the event any better
than this. We have assumed the Markov property such that the transition
probability does not depend on the previous history of the walker;
only the position $x$ prior to making the step. It is normalised
such that
\begin{equation}
\sum_{m=-\infty}^{\infty}T_{n}(x_{m}-x_{m^{\prime}}\vert x_{m^{\prime}})=1,\label{eq:6a}
\end{equation}
 since the total probability that \emph{any} transition is made, starting
from $x_{m^{\prime}}$, is unity. The probability that the walker
is at position $m$ at time $n$ is a sum of probabilities of all
possible previous histories that lead to this situation. In the Markov
case, the master equation shows that these \emph{path} probabilities
are products of transition probabilities and the probability of an
initial situation, a simple viewpoint that we shall exploit later.

\subsection{Kramers-Moyal and Fokker-Planck equations}

The Kramers-Moyal and Fokker-Planck equations describe the evolution
of \emph{probability density functions}, denoted $P$, which are continuous in space
(K-M) and additionally in time (F-P). We start with the Chapman-Kolmogorov
equation, an integral form of the master equation for the evolution
of a probability density function that is continuous in space:
\begin{equation}
P(x,t+\tau)=\int T(\Delta x\vert x-\Delta x,t)P(x-\Delta x,t)d\Delta x.\label{eq:18}
\end{equation}
 We have swapped the discrete time label $n$ for a parameter $t$.
The quantity $T(\Delta x\vert x,t)$ describes a jump from $x$ through
distance $\Delta x$ in a period $\tau$ starting from time $t$.
Note that $T$ now has dimensions of inverse length (it is really
a Markovian transition probability \emph{density}), and is normalised
according to $\int T(\Delta x\vert x,t)d\Delta x=1$.

We can turn this integral equation into a differential equation by
expanding the integrand in $\Delta x$ to get
\begin{equation}
P(x,t+\tau)=P(x,t)+\int d\Delta x\sum_{n=1}^{\infty}\frac{1}{n!}\left(-\Delta x\right)^{n}\frac{\partial^{n}\left(T(\Delta x\vert x,t)P(x,t)\right)}{\partial x^{n}},\label{eq:20}
\end{equation}
 and define the Kramers-Moyal coefficients, proportional to moments
of $T$:
\begin{equation}
M_{n}(x,t)=\frac{1}{\tau}\int d\Delta x(\Delta x)^{n}T(\Delta x\vert x,t),\label{eq:21}
\end{equation}
 to obtain the (discrete time) Kramers-Moyal equation:
\begin{equation}
\frac{1}{\tau}\left(P(x,t+\tau)-P(x,t)\right)=\sum_{n=1}^{\infty}\frac{(-1)^{n}}{n!}\frac{\partial^{n}\left(M_{n}(x,t)P(x,t)\right)}{\partial x^{n}}.\label{eq:22}
\end{equation}
Sometimes the Kramers-Moyal equation is defined with a time derivative
of $P$ on the left hand side instead of a difference.

Equation (\ref{eq:22}) is rather intractable, due to the infinite
number of higher derivatives on the right hand side. However, we might
wish to confine attention to evolution in continuous time, and consider
only stochastic processes which are continuous in space in this limit.
This excludes processes which involve discontinuous jumps: the allowed
step lengths must go to zero as the timestep goes to zero. In this
limit, every coefficient vanishes except the first and second, consistent with the Pawula theorem. Furthermore,
the difference on the left hand side of equation (\ref{eq:22}) becomes
a time derivative and we end up with the Fokker-Planck equation (FPE):
\begin{equation}
\frac{\partial P(x,t)}{\partial t}=-\frac{\partial\left(M_{1}(x,t)P(x,t)\right)}{\partial x}+\frac{1}{2}\frac{\partial^{2}\left(M_{2}(x,t)P(x,t)\right)}{\partial x^{2}}.\label{eq:26}
\end{equation}
We can define a probability current:
\begin{equation}
J=M_{1}(x,t)P(x,t)-\frac{1}{2}\frac{\partial\left(M_{2}(x,t)P(x,t)\right)}{\partial x},\label{eq:27}
\end{equation}
 and view the FPE as a continuity equation for probability density:
\begin{equation}
\frac{\partial P(x,t)}{\partial t}=-\frac{\partial}{\partial x}\left(M_{1}(x,t)P(x,t)-\frac{1}{2}\frac{\partial\left(M_{2}(x,t)P(x,t)\right)}{\partial x}\right)=-\frac{\partial J}{\partial x}.\label{eq:28-1}
\end{equation}
 The FPE reduces to the familiar diffusion equation if we take $M_{1}$
and $M_{2}$ to be zero and $2D$, respectively. Note that it
is probability that is diffusing, not a physical property like gas
concentration. As an example, consider the limit of the symmetric
Markov random walk in one dimension as timestep and spatial step go
to zero: the so-called Wiener process. The probability density $P(x,t)$
evolves according to
\begin{equation}
\frac{\partial P(x,t)}{\partial t}=D\frac{\partial^{2}P(x,t)}{\partial x^{2}},\label{eq:28c}
\end{equation}
 with an initial condition $P(x,0)=\delta(x)$. The statistical properties of the process
are represented by the probability density that satisfies
this equation:
\begin{equation}
P(x,t)=\frac{1}{\left(4\pi Dt\right)^{1/2}}\exp\left(-\frac{x^{2}}{4Dt}\right),\label{eq:33}
\end{equation}
representing the increase in positional uncertainty of the walker
as time progresses.

\subsection{Ornstein-Uhlenbeck process}

\label{ornstein:sec} We now consider a very important stochastic
process describing the evolution of the velocity of a particle. We
shall approach this from a different point of view: a treatment of
the dynamics where Newton's equations are supplemented by environmental
forces, some of which are stochastic. It is proposed that the environment
introduces a linear damping term together with random noise:
\begin{equation}
\dot{v}=-\gamma v+b\xi(t),\label{eq:34}
\end{equation}
 where $\gamma$ is the friction coefficient, $b$ is a constant,
and $\xi$ has statistical properties $\langle\xi(t)\rangle=0$, where
the brackets represent an expectation over the probability distribution
of the noise, and $\langle\xi(t)\xi(t^{\prime})\rangle=\delta(t-t^{\prime})$,
which states that the noise is sampled from a distribution with no
autocorrelation in time. The singular variance of the noise might
seem to present a problem but this can be accommodated. This is the Langevin
equation. We can demonstrate that it is equivalent to a description
based on a Fokker-Planck equation by evaluating the K-M coefficients,
considering equation (\ref{eq:21}) in the form
\begin{equation}
M_{n}(v,t)=\frac{1}{\tau}\int d\Delta v(\Delta v)^{n}T(\Delta v\vert v,t)=\frac{1}{\tau}\langle\left(v(t+\tau)-v(t)\right)^{n}\rangle,\label{eq:37}
\end{equation}
 and in the continuum limit where $\tau\rightarrow0$. This requires an equivalence
between the average of $\left(\Delta v\right)^{n}$ over a transition
probability density $T$, and the average over the statistics of the
noise $\xi$. We integrate equation (\ref{eq:34}) for small $\tau$
to get
\begin{equation}
v(t+\tau)-v(t)=-\gamma\int_{t}^{t+\tau}vdt+b\int_{t}^{t+\tau}\xi(t^{\prime})dt^{\prime}\approx-\gamma v(t)\tau+b\int_{t}^{t+\tau}\xi(t^{\prime})dt^{\prime},\label{eq:38}
\end{equation}
 and according to the properties of the noise and in the limit $\tau\rightarrow0$
this gives $\langle dv\rangle=-\gamma v\tau$ with $dv=v(t+\tau)-v(t)$,
such that $M_{1}(v)=\langle\dot{v}\rangle=-\gamma v$. We also construct
$\left(v(t+\tau)-v(t)\right)^{2}$ and using the appropriate statistical
properties and the continuum limit, we get $\langle\left(dv\right)^{2}\rangle=b^{2}\tau$
and $M_{2}=b^{2}$. We have therefore established that the FPE equivalent
to the Langevin equation (\ref{eq:34}) is
\begin{equation}
\frac{\partial P(v,t)}{\partial t}=\frac{\partial\left(\gamma vP(v,t)\right)}{\partial v}+\frac{b^{2}}{2}\frac{\partial^{2}P(v,t)}{\partial v^{2}}.\label{eq:40}
\end{equation}
The stationary solution to this equation ought to be the Maxwell-Boltzmann
velocity distribution $P(v)\propto\exp\left(-mv^{2}/2k_{B}T\right)$
of a particle of mass $m$ in thermal equilibrium with a bath at temperature
$T$, so $b$ must be related to $T$ and $\gamma$ in the form $b^{2}=2k_{B}T\gamma/m$,
where $k_{B}$ is Boltzmann's constant. This is a connection known
as a fluctuation dissipation relation: $b$ characterises the fluctuations
and $\gamma$ the dissipation or damping in the Langevin equation. Furthermore,
it may be shown that the time-dependent solution to equation (\ref{eq:40}),
with initial condition $\delta(v-v_{0})$ at time $t_{0}$, is
\begin{equation}
P_{\mathrm{OU}}^{T}\left[v,t|v_{0},t_{0}\right]=\sqrt{\frac{m}{2\pi k_{B}T(1-{\rm e}^{-2\gamma(t-t_{0})})}}\exp\left(-\frac{m\big(v-v_{0}{\rm e}^{-\gamma(t-t_{0})}\big)^{2}}{2k_{B}T\left(1-{\rm e}^{-2\gamma(t-t_{0})}\right)}\right).\label{eq:41}
\end{equation}
 This is a gaussian with time-dependent mean and variance. The notation
$P_{\mathrm{OU}}^{T}[\cdots]$ is used to denote a transition probability
density for this so-called Ornstein-Uhlenbeck process starting from
initial value $v_{0}$ at initial time $t_{0}$, and ending at the
final value $v$ at time $t$.

The same mathematics can be used to describe the motion of a particle
in a harmonic potential $\phi(x)=\kappa x^{2}/2$, in the limit that
the frictional damping coefficient $\gamma$ is very large. The Langevin
equations that describe the dynamics are $\dot{v}=-\gamma v-\kappa x/m+b\xi(t)$
and $\dot{x}=v,$ which reduce in this so-called overdamped limit
to
\begin{equation}
\dot{x}=-\frac{\kappa}{m\gamma}x+\frac{b}{\gamma}\xi(t),\label{eq:43}
\end{equation}
 which then has the same form as equation (\ref{eq:34}), but for
position instead of velocity. The transition probability (\ref{eq:41}),
recast in terms of $x$, therefore can be employed.

In summary, the evolution of a system interacting with a coarse-grained
environment can be modelled using a stochastic treatment that includes time-dependent
random external forces. However, these really represent the effect
of uncertainty in the \emph{initial} conditions for the system and
its environment: indefiniteness in some of those initial environmental
conditions might only have an impact upon the system at a later time.
For example, the uncertainty in the velocity of a particle in a gas
increases as particles that were initially far away, and that were
poorly specified at the initial time, have the opportunity to move
closer and interact. The evolution equations are not time reversal
symmetric since the principle of causality is assumed: the probability
of a system configuration depends upon events that precede it in time,
and not on events in the future. The evolving probability density
can capture the growth in configurational uncertainty with time. We
can now explore how growth of uncertainty in system configuration
might be related to entropy production and the irreversibility of
macroscopic processes.

\section{Entropy generation and stochastic irreversibility}

\subsection{The reversibility of a stochastic trajectory}

The usual statement of the second law in thermodynamics is that it
is impossible to observe the reverse of an entropy producing process.
Let us immediately reject this version of the law and recognise that
nothing is impossible. A ball might roll off a table and land at our
feet. But there is never stillness at the microscopic level and, without
breaking any law of mechanics, the molecular motion of the air, ground
and ball might conspire to reverse their macroscopic motion, bringing
the ball back to rest on the table. This is not ridiculous: it is
an inevitable consequence of the time reversal symmetry of Newton's
laws. All we need for this event to occur is to create the right initial
conditions. Of course, that is where the problem lies: it is virtually
impossible to engineer such a situation, but virtually impossible
is not absolutely impossible.

This of course highlights the point behind Loschmidt's paradox. If we were to time
reverse the equations of motion of every atom that was involved in
the motion of the ball at the end of such an event we \emph{would} observe the
reverse behaviour. Or rather more suggestively, we would observe both
the forward \emph{and} the reverse behaviour with probability 1. This
of course is such an overwhelmingly difficult task that one would
never entertain the idea of its realisation. Indeed it is also not
how one typically considers irreversibility in the real world, whether
that be in the lab or through experience. What one might in principle
be able to investigate is the explicit time reversal of just
the motion of the particle(s) of interest to see whether the previous history can be reversed. Instead of reversing
the motion of all the atoms of the ground, the air etc, we just attempt
to roll the ball back towards the table at the same speed at which
it landed at our feet. In this scenario we certainly would not expect
the reverse behaviour. Now because the reverse motion is not inevitable
we have somehow, for the system we are considering, identified (or
perhaps constructed) the concept of irreversibility albeit on a somewhat anthropic
level: events do not easily run backwards. How have we evaded Loschmidt's paradox here? We failed to provide the initial conditions that would ensure reversibility: we left out the reversal of the motion of all the
other atoms. If they act upon the system differently under time reversal then irreversibility is (virtually) inevitable.
This is not so very profound, but what we
have highlighted here is the one of the principle paradigms of
thermodynamics, the separation of the system of interest and its environment,
or for our example the ball and the rest of the surroundings. Given
then that we expect such irreversible behaviour when we ignore the details of
the environment in this way, we can ask what representation of that environment might be most suitable when establishing a measure of the irreversibility of the process?
The answer to which is when the environment explicitly
interacts with the system in such a way that time reversal is irrelevant.
Whilst never strictly true, this can hold as a limiting case which can be represented in a model, allowing us to determine the extent to which the reversal of just the velocities of the system components can lead to a retracing of the previous sequence of events. Stochastic dynamics can provide an example of such
a model. In the appropriate limits, we may consider the collective
influence of all the atoms in the environment to act on the system in the same inherently
unpredictable and dissipative way regardless of whether their coordinates are time reversed or not. In the Langevin equation this is achieved by
ignoring a quite startling number of degrees of freedom associated
with the environment, idealising their behaviour as noise along with
a frictional force which slows the particle regardless of which way
it is travelling. If we consider now the motion of our system of interest
according to this Langevin scheme both its forward and reverse motion are no
longer certain and we can attribute a probability to each path under the
influence of the environmental effects. How might we measure irreversibility
given these dynamics? It is no longer appropriate to consider whether
upon time reversal the exact path is retraced since the paths are
stochastic. Indeed in the continuous limit the probability of this
happening tends to zero. So we ask the question, what is the probability
of observing some forward process compared to the probability of seeing
that forward process undone? Or perhaps, to what extent has the introduction
of stochastic behaviour violated Loschmidt's expectation? This section
is largely devoted to the formulation of such a quantity.\\
 \\
 Intuitively we understand that we should be comparing the probability
of observing some forward and reverse behaviour, but these ideas
need to be made concrete. Let us proceed in a manner that allows us
to make a more direct connection between irreversibility and our consideration
of Loschmidt's paradox. First, let us imagine a system which evolves
under some suitable stochastic dynamics. We specifically
consider a realisation
or trajectory that runs from time $t=0$ to $t=\tau$. Throughout
this process we imagine that any number of system parameters may be
subject to change. This could be, for example under suitable Langevin
dynamics, the temperature of the heat bath or perhaps the nature of
a confining potential. The effect of these changes in the parameters
is to alter the probabilistic behaviour of the system as time evolves.
Following the literature we assume that any such change in these system
parameters occurs according to some protocol $\lambda(t)$ which itself
is a function of time. A particular realisation is not guaranteed
to take place, since the system is stochastic, so consequently we
associate with it a probability of occurring which will be entirely
dependent on the exact trajectory taken, for example $x(t)$, and
the protocol $\lambda(t)$. \\
 \\
We can readily compare different probabilities
associated with different paths and protocols. To quantify an irreversibility
in the sense of the breaking of Loschmidt's expectation however, we must consider one specific path
and protocol. Recall now our definition of the paradox. In a deterministic
system, a time reversal of all the variables at the end of a process
of length $\tau$ leads to the observation of the reverse behaviour
with probability 1 over the same period $\tau$. It is the probability
of the trajectory which corresponds to this reverse behaviour within
a stochastic system that we must address. To do so let us consider
what we mean by time reversal. A time reversal can be thought of as
the operation of the time reversal operator, $\hat{T}$ on the system
variables and distribution. Specifically for position $x$, momentum
$p$ and some protocol $\lambda$ we have $\hat{T}x=x$, $\hat{T}p=-p$
and $\hat{T}\lambda=\lambda$. If we were to do this after time $\tau$
for a set of Hamilton's equations of motion in which the protocol
was time-independent, the trajectory would be the exact time reversed
retracing of the forward trajectory. We shall call this trajectory
the \emph{reversed trajectory} and is phenomenologically the `running
backwards' of the forward behaviour. Similarly, if we were to consider
a motion in a deterministic system that was subject to some protocol
(controlling perhaps some external field), we would observe the reversed
trajectory only
if the original protocol were performed symmetrically backwards. This
running of the protocol backwards we shall call the \emph{reversed
protocol}. We now are in a position to construct a measure of irreversibility
in a stochastic system. We do so by comparing the probability of observing
the forward trajectory under the forward protocol with the probability
of observing the reversed trajectory under the reversed protocol following
a time reversal at the end of the forward process. We literally attempt
to undo the forward process and measure how likely that is. Since
the quantities we have just defined here are crucial to this chapter
we spend a moment making their nature absolutely clear before we proceed.
To reiterate we wish to consider a:
\begin{itemize}
\item Reversed trajectory:\\
 Given a trajectory $X(t)$ that runs from time $t=0$ to $t=\tau$,
we define the reversed trajectory $\bar{X}(t)$ which runs forwards
in time explicitly such that $\bar{X}(t)=\hat{T}X(\tau-t)$. Examples
are for position $\bar{x}(t)=x(\tau-t)$ and for momentum $\bar{p}(t)=-p(\tau-t)$.
\item Reversed protocol:\\
 The protocol $\lambda(t)$ behaves in the same way as the position
variable $x$ under time reversal and so we define the reversed
protocol $\bar{\lambda}(t)$ such that $\bar{\lambda}(t)=\lambda(\tau-t)$
\end{itemize}
Given these definitions we can construct the path probabilities we
seek to compare. For notational clarity we label path probabilities
that depend upon the forward protocol $\lambda(t)$ with the superscript
$F$ to denote the forward process and probabilities which depend
upon the reversed protocol $\bar{\lambda}(t)$ with the superscript
$R$ to denote the reverse process. The probability of observing
a given trajectory $X$, $\mathcal{P}^{F}[X]$, has two components. First is
the probability of the path given its starting point $X(0)$ which
we shall write as $\mathcal{P}^{F}[X(\tau)|X(0)]$ and second is the initial
probability of being at the start of the path, which we write as $\mathcal{P}_{{\rm start}}(X(0))$
since it concerns the distribution of variables at the start of the
forward process. The probability of observing the forward path is
then given as
\begin{equation}
\mathcal{P}^{F}[X]=\mathcal{P}_{{\rm start}}(X(0))\mathcal{P}^{F}[X(\tau)|X(0)].
\end{equation}
 We can proceed along these lines; however it is often more intuitive
to proceed if we imagine the path probability as being approximated
by a sequence of jumps that occur at distinct times. Since continuous
stochastic behaviour can be readily approximated by jump processes,
but not the other way round, this simultaneously allows us to generalise
any statements for a wider class of Markov processes. We shall assume
for brevity that the jump processes occur in discrete time; however
we note that the final quantity we shall arrive at would be identical
if such a constraint were relaxed (we point the interested reader
to \cite{adiabaticnonadiabatic0}, for example). By repeated application
of the Markov property for such a system we can write
\begin{align}
\mathcal{P}^{F}[X] & =\mathcal{P}_{{\rm start}}(X_{0})\mathcal{P}(X_{1}|X_{0},\lambda(t_{1}))\times \mathcal{P}(X_{2}|X_{1},\lambda(t_{2}))\times\ldots\nonumber \\
 & \qquad\ldots\times \mathcal{P}(X_{n}|X_{n-1},\lambda(t_{n})).
\end{align}
 Here we consider a trajectory that is approximated by the jump sequence
between $n+1$ points $X_{0}$,$X_{1}$, $\ldots X_{n}$ such that
there are $n$ distinct transitions which occur at discrete times
$t_{1},t_{2},\ldots t_{n}$, and where $X_{0}=X(0)$ and $X_{n}=X(\tau)$.
$\mathcal{P}(X_{i}|X_{i-1},\lambda(t_{i}))$ is the probability of a jump from
$X_{i-1}$ to $X_{i}$ using the value of the protocol evaluated at
time $t_{i}$.

\begin{figure}[!htp]
\begin{centering}
\includegraphics[clip,width=100mm]{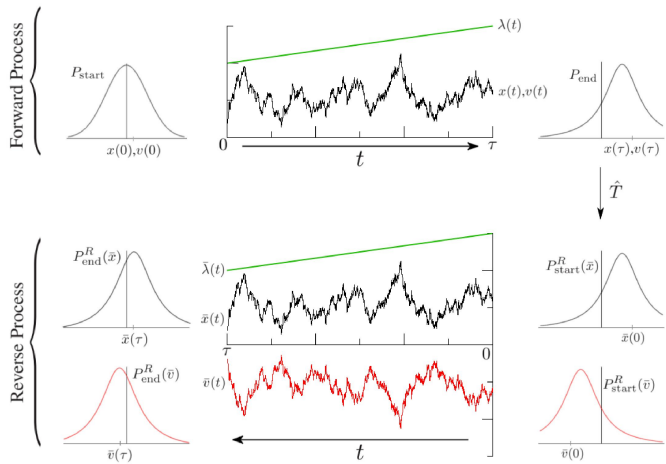} \caption{An illustration of the definition of the forward and reverse processes.
The forward process consists of an initial probability \emph{density}, $P_{{\rm start}}$ which evolves forward in time under the forward protocol $\lambda(t)$ over a period $\tau$, at the end of which the variable is distributed according to $P_{{\rm end}}$. The reverse process consists of evolution from the distribution $P^R_{{\rm start}}$, which is related to $P_{{\rm end}}$ by a time reversal, under the reversed protocol $\bar{\lambda}(t)$ over the same period $\tau$, at the end of which the system will be distributed according to some final distribution $P^R_{{\rm end}}$, which in general is not related to $P_{{\rm start}}$ and does not explicitly feature in assessment of the irreversibility of the forward process. A particular realisation of the forward process is characterised by the forward trajectory $X(t)$,  illustrated here as being $x(t)$ or $v(t)$. To determine the irreversibility of this realisation, the reversed trajectories, $\bar{x}(t)$ or $\bar{v}(t)$, related by a time reversal, need to be considered as realisations in the reverse process.
}
\label{forback}
\end{centering}

\end{figure}

Continuing with our description of irreversibility we construct the
probability of the reversed trajectory under the reversed protocol.
Approximating as a sequence of jumps as before we may write
\begin{align}
\mathcal{P}^{R}[\bar{X}] & =\hat{T}\mathcal{P}_{{\rm end}}(\bar{X}(0))\mathcal{P}^{R}[\bar{X}(\tau)|\bar{X}(0)]\nonumber \\
 & =\mathcal{P}_{{\rm start}}^{R}(\bar{X}(0))\mathcal{P}^{R}[\bar{X}(\tau)|\bar{X}(0)]\nonumber \\
 & =\mathcal{P}_{{\rm start}}^{R}(\bar{X}_{0})\mathcal{P}(\bar{X}_{1}|\bar{X}_{0},\bar{\lambda}(t_{1}))\times\ldots\times \mathcal{P}(\bar{X}_{n}|\bar{X}_{n-1},\bar{\lambda}(t_{n}))
\end{align}
 There are two key concepts here. The first, in accordance with our
definition of irreversibility, is that we attempt to `undo' the motion
from the end of the forward process and so the initial distribution
is formed from the distribution to which $\mathcal{P}_{{\rm start}}$ evolves
under $\lambda(t)$, such that for continuous probability density distributions we have
\begin{equation}
P_{{\rm end}}(X(\tau))=\int dX\; P_{{\rm start}}(X(0))P^{F}[X(\tau)|X(0)],
\end{equation}
 so named because it is the probability distribution at the end of
the forward process. For our discrete model the equivalent is given
by
\begin{equation}
\mathcal{P}_{{\rm end}}(X_{n})=\sum_{X_0}\ldots\sum_{X_{n-1}}\prod_{i=0}^{n-1}\mathcal{P}(X_{i+1}|X_{i},\lambda(t_{i+1}))\mathcal{P}_{{\rm start}}(X_{0}).
\end{equation}
 Secondly, to attempt to observe the reverse trajectory starting from
$X(\tau)$ we must perform a time reversal of our system to take advantage of
the reversibility in Hamilton's equations. However, when we time reverse
the variable $X$ we are obliged to transform the distribution $\mathcal{P}_{{\rm end}}$
as well, since the likelihood of starting the reverse trajectory with variable $\hat{T}X$
after we time reverse $X$ is required to be the same as the likelihood of arriving
at $X$ before the time reversal. This transformed $\mathcal{P}_{{\rm end}}$,
$\hat{T}\mathcal{P}_{{\rm end}}$, is the initial distribution for the reverse
process and is thus labelled $\mathcal{P}_{{\rm start}}^{R}$. Analogously,
evolution under $\bar{\lambda}(t)$ takes the system distribution
to $\mathcal{P}_{\mathrm{end}}^{R}$. The forward process and its relation to
the reverse process are illustrated for both coordinates $x$ and
$v$, which do and do not change sign following time reversal, respectively,
in Figure \ref{forback}, along with illustrations of the reversed
trajectories and protocols.

Let us now form our prototypical measure of the irreversibility of
the path $X$, which for now we denote $I$:
\begin{equation}
I[X]=\ln{\left[\frac{\mathcal{P}^{F}[X]}{\mathcal{P}^{R}[\bar{X}]}\right]}
\end{equation}
 There are some key points to notice about such a quantity. First,
since $\bar{X}$ and $X$ are simply related, $I$ is a functional
of the trajectory $X$ and accordingly will take a range of values over
all the possible `realisations' of the dynamics: as such it will be
characterised by a probability distribution. Further, there is nothing
in its form which disallows negative values. Finally the quantity
vanishes if the reversed trajectory occurs with the same probability
as the forward trajectory under the relevant protocols: a process
is deemed reversible if the forward process can be `undone' with equal
probability. We can simplify this form since we know how the time
reversed protocols and trajectories are related. Given the step sequence
laid out for the approximation to a continuous trajectory we can transform
$X$ and $t$ according to $\bar{X}_{i}=\hat{T}X_{n-i}$ and $\bar{\lambda}(t_{i})=\lambda(t_{n-i+1})$
giving
\begin{align}
\mathcal{P}^{R}[\bar{X}] & =\mathcal{P}_{{\rm start}}^{R}(\hat{T}X(\tau))\mathcal{P}^{R}[\hat{T}{X}(0)|\hat{T}X(\tau)]\nonumber \\
 & =\mathcal{P}_{{\rm start}}^{R}(\hat{T}X_{n})\mathcal{P}(\hat{T}X_{n-1}|\hat{T}{X}_{n},{\lambda}(t_{n}))\times\ldots\times \mathcal{P}(\hat{T}{X}_{0}|\hat{T}{X}_{1},{\lambda}(t_{1}))
\end{align}
 Pointing out that $\mathcal{P}_{{\rm start}}^{R}(\hat{T}X_{n})=\hat{T}\mathcal{P}_{{\rm end}}(\hat{T}X_{n})=\mathcal{P}_{{\rm end}}(X_{n})$
we thus have
\begin{align}
\ln{\left[\frac{\mathcal{P}^{F}[X]}{\mathcal{P}^{R}[\bar{X}]}\right]} & =\ln{\left(\frac{\mathcal{P}_{{\rm start}}(X(0))}{\mathcal{P}_{{\rm end}}(X(\tau))}\right)}+\ln{\left[\frac{\mathcal{P}^{F}[X(\tau)|X(0)]}{\mathcal{P}^{R}[\hat{T}X(0)|\hat{T}X(\tau)]}\right]}\nonumber \\
 & =\ln{\left[\frac{\mathcal{P}_{{\rm start}}(X_{0})}{\mathcal{P}_{{\rm end}}(X_{n})}\prod_{i=1}^{n}\frac{\mathcal{P}(X_{i}|X_{i-1},\lambda(t_{i}))}{\mathcal{P}(\hat{T}X_{i-1}|\hat{T}X_{i},\lambda(t_{i}))}\right]}\label{ratio}
\end{align}
 Let us explicitly consider this quantity for a specific model to
understand its meaning in physical terms. Consider the continuous stochastic process described by
the Langevin equation from section \ref{ornstein:sec}, where $X=v$ and we have
\begin{equation}
\dot{v}=-\gamma v+\left(\frac{2k_{B}T(t)\gamma}{m}\right)^{1/2}\xi(t),
\end{equation}
 where $\xi(t)$ is white noise. The equivalent Fokker-Planck equation
is given by
\begin{equation}
\frac{\partial P(v,t)}{\partial t}=\frac{\partial\left(\gamma vP(v,t)\right)}{\partial v}+\frac{k_{B}T(t)\gamma}{m}\frac{\partial^{2}P(v,t)}{\partial v^{2}}.\label{FPV}
\end{equation}
 where $P$ is a probability density. By insertion of probability densities and infinitesimal volumes into equation (\ref{ratio}) and cancelling the latter we observe that we may use probability densities to represent the quantity $I[X]$ for this continuous behaviour without a loss of generality. To introduce a distinct forward and reverse process, let us allow
the temperature to vary with a protocol $\lambda(t)$. We choose for
simplicity a protocol which consists only of step changes such that
\begin{equation}
T(\lambda(t_{i}))=T_{j}\quad t_{i}\in[(j\!-\!1)\Delta t,j\Delta t],
\end{equation}
 where $j$ is an integer in the range $1\leq j\leq N$, such that
$N\Delta t=\tau$. Because the process is simply the combination of different
Ornstein-Uhlenbeck processes each of which is characterised by defined solution equation
(\ref{eq:41}) we can represent the path probability in a piecewise fashion.
Consolidating with our notation, the continuous Langevin behaviour
at some fixed temperature can be considered to be the limit, $dt=(t_{i+1}-t_{i})\to0$,
of the discrete jump process, so that
\begin{align}
 & \lim_{dt\to0}\prod_{t_{i}=(j-1)\Delta t}^{t_{i}=j\Delta t}\! \mathcal{P}(v_{i}|v_{i-1},\lambda(t_{i}))=P_{{\rm OU}}^{T_{j}}[v(j\Delta t)|v((j-1)\Delta t)]dv(j\Delta t)=\nonumber \\
 & \left(\frac{m}{2\pi k_{B}T_{j}(1-{\mathrm{e}}^{-2\gamma\Delta t})}\right)^{1/2}\exp\left(-\frac{m\big(v(j\Delta t)-v((j-1)\Delta t){\mathrm{e}}^{-\gamma\Delta t}\big)^{2}}{2k_{B}T_{j}\left(1-{\mathrm{e}}^{-2\gamma\Delta t}\right)}\right)dv(j\Delta t).
\end{align}
 The total conditional path probability density (with units equal to the inverse dimensionality of the path) over $N$ of these step changes in temperature is then
by application of the Markov property
\begin{align}
 & P^{F}[v(\tau)|v(0)]=\nonumber \\
 & \prod_{j=1}^{N}\left(\frac{m}{2\pi k_{B}T_{j}(1-{\mathrm{e}}^{-2\gamma\Delta t})}\right)^{1/2}\exp\left(-\frac{m\big(v(j\Delta t)-v((j-1)\Delta t){\mathrm{e}}^{-\gamma\Delta t}\big)^{2}}{2k_{B}T_{j}\left(1-{\mathrm{e}}^{-2\gamma\Delta t}\right)}\right)
\end{align}
 and since $\hat{T}v=-v$
\begin{align}
 & P^{R}[-v(0)|-v(\tau)]=\nonumber \\
 & \prod_{j=1}^{N}\left(\frac{m}{2\pi k_{B}T_{j}(1-\mathrm{e}^{-2\gamma\Delta t})}\right)^{1/2}\exp\left(-\frac{m\big(-v((j-1)\Delta t)+v(j\Delta t)\mathrm{e}^{-\gamma\Delta t}\big)^{2}}{2k_{B}T_{j}\left(1-\mathrm{e}^{-2\gamma\Delta t}\right)}\right)
\end{align}
 Taking the logarithm of their ratio explicitly and abbreviating $v(j\Delta t)=v_{j}$
yields
\begin{equation}
\ln{\left[\frac{P^{F}[v(\tau)|v(0)]}{P^{R}[-v(0)|-v(\tau)]}\right]}=-\frac{1}{k_B}\sum_{j=1}^{N}\frac{m}{2T_{j}}\left(v_{j}^{2}-v_{{j\!-\!1}}^{2}\right)\label{heat}
\end{equation}
 which is quite manifestly equal to the sum of negative changes of
the kinetic energy of the particle scaled by $k_B$  and the environmental temperature to which
the particle is exposed. Our model consists only of the particle and
the environment and so each negative kinetic energy change of the particle, $-\Delta Q$, must be associated
with a positive flow of heat $\Delta Q_{\rm med}$ into the environment such that we define $\Delta Q_{\rm med}=-\Delta Q$. For the Langevin
equation the effect of the environment is idealised as a dissipative
friction term and a fluctuating white noise characterised by a defined
temperature which is entirely independent of the behaviour of the
particle. This is the idealisation of a large equilibrium heat bath
for which the exchanged heat is directly related to the entropy change
of the bath through the relation $\Delta Q_{\rm med}=T\Delta S$. It may be
argued that changing between $N$ temperatures under such an idealisation
is equivalent to exposing the particle to $N$ separate equilibrium
baths each experiencing an entropy change according to $\Delta Q_{{\rm med},j}=T_{j}\Delta S_{j}$.
We consequently assert, for this particular model at least, that
\begin{equation}
k_{B}\ln{\left[\frac{P^{F}[v(\tau)|v(0)]}{P^{R}[-v(0)|-v(\tau)]}\right]}=\sum_{j}\frac{\Delta Q_{{\rm med},j}}{T_{j}}=\sum_{j}\Delta S_{j}=\Delta S_{\rm{med}}
\end{equation}
 where the entropy production in all $N$ baths can be denoted as a total
entropy production $\Delta S_{\mathrm{med}}$ that occurs in a generalised medium. \\
\\
Let us now examine the remaining part of our quantification of irreversibility
which here is given in equation (\ref{ratio}) by the logarithm of the ratio of $P_{{\rm start}}(v(0))$
and $P_{{\rm end}}(v(\tau))$. Given an arbitrary initial distribution
one can write this as the change in the logarithm of the dynamical
solution to $P$ as given by the Fokker-Planck equation (\ref{FPV}).
Consequently we can write
\begin{equation}
\ln{\left(\frac{P_{{\rm start}}(v(0))}{P_{{\rm end}}(v(\tau))}\right)}=\ln{\frac{P(v,0)}{P(v,\tau)}}=-\left(\ln{P(v,\tau)}-\ln{P(v,0)}\right).
\end{equation}
 If we now characterise the \emph{mean} entropy of our Langevin particle
or `system' with a Gibbs entropy which we allow to be time dependent
such that
\begin{equation}
\langle S_{{\rm sys}}\rangle=S_{\mathrm{Gibbs}}=-k_{B}\int dv\; P(v,t)\ln{P(v,t)}
\label{meansys1}
\end{equation}
 one can make the conceptual leap that it is an individual value for the
entropy of the system for a given $v$ and time $t$ that is
being averaged in the above integral \cite{seifertoriginal}%
\footnote{Strictly $P(v,t)$ is a probability density and so for equation (\ref{meansys1}) to be consistent
with the entropy arising from the combinatoric arguments of statistical
mechanics and dimensionally correct it might be argued we should be
considering $\ln\left(P(v,t)dv\right)$. However, for relative changes this issue is irrelevant.%
}, $S_{{\rm sys}}=-k_{B}\ln{P(v,t)}$. If we accept these assertions
we find that our measure of irreversibility for any one individual
trajectory is formed as
\begin{equation}
k_{B}I[X]=\Delta S_{{\rm sys}}+\Delta S_{{\rm med}}.
\end{equation}
 Since our model consists only of the Langevin particle (the system)
and a heat bath (the medium) we therefore regard this sum as
the total entropy production associated with such a trajectory and
make the assertion that our measure of irreversibility is identically
the increase in the total entropy of the universe
\begin{equation}
\Delta S_{{\rm tot}}[X]=\Delta S_{{\rm sys}}+\Delta S_{{\rm med}}=k_{B}\ln{\left[\frac{P^{F}[X]}{P^{R}[\bar{X}]}\right]}
\label{totalent}
\end{equation}
 in this model at least. However, we have already stated that nothing
prevents this quantity from taking negative values. If this is to
be the total entropy production, how is this permitted given our knowledge
of the second law of thermodynamics? In essence, describing the way
in which a quantity that looks like the total entropy production can take both
positive and negative values, but obeys well defined statistical requirements
such that, for example, it is compatible with the second law, is the
subject matter of the so-called fluctuation theorems or fluctuation
relations. These relations are disarmingly simple, but allow us to
make predictions far beyond those possible in classical thermodynamics.
For this class of system in fact, they are so simple we can derive
in a couple of lines a most fundamental relation and immediately reconcile
the second law in terms of our irreversibility functional. Let us
consider the average, with respect to all possible forward realisations,
of the quantity $\exp{(-\Delta S_{{\rm tot}}[X]/k_{B})}$ which we
write $\langle\exp{(-\Delta S_{{\rm tot}}[X]/k_{B})}\rangle$ and
where the angled brackets denote a weighted path integration. Performing
the average yields
\begin{align}
\langle{\rm e}^{-\Delta S_{{\rm tot}}[X]/k_{B}}\rangle & =\int dX\; P^{F}[X]{\rm e}^{-\Delta S_{{\rm tot}}[X]/k_{B}}\nonumber \\
 & =\int dX\; P^{F}[X]\frac{P^{R}[\bar{X}]}{P^{F}[X]}\nonumber \\
 & =\int d\bar{X}P^{R}[\bar{X}]
 \label{IFTdev}
\end{align}
 where we assume the measures for the path integrals are related by
$dX=d\bar{X}$ since the reverse trajectory is defined on the same
space. Or perhaps more transparently, in the discrete approximation
multiple summations over $X_{0},\dots X_{n}$ yield the same result
as summation over $X_{n},\dots X_{0}$. The expression above now trivially
integrates to unity which allows us to write a so-called \cite{seifertoriginal}\\
 \\
 \emph{Integral Fluctuation Theorem}
\begin{equation}
\langle{\rm e}^{-\Delta S_{{\rm tot}}[X]/k_{B}}\rangle=1
\end{equation}
 This remarkably simple relation holds for all times, protocols and
initial conditions%
\footnote{We do though assume that nowhere in the initial available configuration space do we have $P_{{\rm start}}(X)=0$. This is
a paraphrasing of the so-called ergodic consistency requirement found in deterministic systems \cite{Evans02} and insists that there must be a trajectory for every possible reversed trajectory and vice versa, so that the all possible paths, $\bar{X}(t)$, are included in the integral in the final line of equation (\ref{IFTdev}).} and implies that the possibility of negative total entropy change
is obligatory. Further, if we make use of Jensen's inequality
\begin{equation}
\langle\exp{(z)}\rangle\geq\exp{\langle z\rangle}
\end{equation}
 we can directly infer
\begin{equation}
\langle\Delta S_{{\rm tot}}\rangle\geq0.
\end{equation}
 Since this holds for any initial condition we may also state that
the mean total entropy monotonically increases for any process. This
statement, under the stochastic dynamics we consider, is the second
law. It is a replacement of, or reinterpretation of equation (\ref{eq:5}). The mean entropy production rate is always positive, but not
necessarily in detail for individual realisations. The second law,
when correctly understood, is statistical in nature and we have now
obtained an expression which places a fundamental bound on those statistics.

\section{Entropy Production in the Overdamped Limit}

We have formulated a quantity which we assert to be the total entropy
production, though it is for a very specific system and importantly
has no ability to describe the application of work. To broaden the
scope of application it is instructive to obtain a general expression
like that obtained in equation (\ref{heat}), but for a class of stochastic
behaviour where we can formulate and verify the total entropy
production without the need for an exact analytical result. This is straightforward for systems with detailed balance \cite{Crooks98}, however we can generalise further. The class
of stochastic behaviour we shall consider will be the simple overdamped
Langevin equation that we discussed in section \ref{ornstein:sec}
involving a position variable described by
\begin{equation}
\dot{x}=\frac{{\cal F}(x)}{m\gamma}+\left(\frac{2k_{B}T}{m\gamma}\right)^{1/2}\xi(t),
\end{equation}
 along with an equivalent Fokker-Planck equation
\begin{equation}
\frac{\partial P(x,t)}{\partial t}=-\frac{1}{m\gamma}\frac{\partial\left({\cal F}(x)P(x,t)\right)}{\partial x}+\frac{k_{B}T}{m\gamma}\frac{\partial^{2}P(x,t)}{\partial x^{2}}.
\end{equation}
 The description includes a force term ${\cal F}(x)$ which allows us to model most
simple thermodynamic processes including the application of work.
We describe the force as a sum of two contributions which
arise respectively from a potential $\phi(x)$ and an external force $f(x)$ which
is applied directly to the particle, both of which we allow to vary
in time through application of a protocol such that
\begin{equation}
{\cal F}(x,\lambda_{0}(t),\lambda_{1}(t))=-\frac{\partial\phi(x,\lambda_{0}(t))}{\partial x}+f(x,\lambda_{1}(t)).
\end{equation}
 The first step in characterising the entropy produced in the medium
according to this description is to identify the main thermodynamic
quantities including the heat exchanged with the bath. To do this
we paraphrase Sekimoto and Seifert  \cite{sekimoto1,sekimoto2,seifertoriginal} and start from basic thermodynamics
and the first law
\begin{equation}
\Delta E=\Delta Q+\Delta W
\end{equation}
 which must hold rigorously despite the stochastic nature of our model.
To proceed, let us consider the change in each of these quantities
in response to evolving our system by a small time $dt$ and corresponding
displacement $dx$. We can readily identify that the system energy
for overdamped conditions is equal to the value of the conservative
potential such that
\begin{equation}
dE=dQ+dW=d(\phi(x,\lambda_{0}(t))).
\end{equation}
 However, at this point we reach a subtlety in the mathematics originating
in the stochastic nature of $x$. Where normally we could describe
the small change in $\phi$ using the usual chain rule of calculus,
when $\phi$ is a function of the stochastic variable $x$ we must
be more careful. The peculiarity is manifest in an ambiguity of expressing
the multiplication of a  continuous stochastic function by a stochastic increment.
The product, which strictly should be regarded as a stochastic integral,
is not uniquely defined because both function and increment cannot
be assumed to behave smoothly on any timescale. The mathematical details \cite{Gardiner09}
are not of our concern for this chapter and so we shall not rigorously
discuss stochastic calculus, or go beyond the following steps of reasoning
and assumption. First we assume that in order to work with thermodynamic
quantities in the traditional sense, as in undergraduate physics,
we require a small change to resemble that of normal calculus, and
this requires, in all instances, multiplication to follow so-called
Stratonovich rules. These rules, denoted in this chapter by the symbol
$\circ$, are taken to mean evaluation of the preceding stochastic
function at the mid-point of the following increment. Following this
procedure we may write
\begin{align}
dE & =d(\phi(x(t),\lambda_{0}(t)))\nonumber \\
 & =\frac{\partial\phi(x(t),\lambda_{0}(t))}{\partial\lambda_{0}}\frac{d\lambda_{0}(t)}{dt}dt+\frac{\partial\phi(x(t),\lambda_{0}(t))}{\partial x}\circ dx\label{eq:54}
\end{align}
Next we can explicitly write down the work from basic mechanics
as contributions from the change in potential and the operation of
an external force:
\begin{equation}
dW=\frac{\partial\phi(x(t),\lambda_{0}(t))}{\partial\lambda_{0}}\frac{d\lambda_{0}(t)}{dt}dt+f(x(t),\lambda_{1}(t))\circ dx.
\end{equation}
 Accordingly we directly have an expression for the heat transfer
to the system in response to a small change $dx$
\begin{align}
dQ & =\frac{\partial\phi(x(t),\lambda_{0}(t))}{\partial x}\circ dx-f(x(t),\lambda_{1}(t))\circ dx\nonumber \\
 & =-{\cal F}(x(t),\lambda_{0}(t),\lambda_{1}(t))\circ dx.
\label{smallheat}
\end{align}
 We may then integrate these small increments over a trajectory of
duration $\tau$ to find
\begin{align}
\Delta E & =\int_{0}^{\tau}dE=\int_{0}^{\tau}d(\phi(x(t),\lambda_{0}(t)))=\phi(x(\tau),\lambda_{0}(\tau))-\phi(x(0),\lambda_{0}(0))\nonumber \\
 & =\Delta\phi
\end{align}

\begin{equation}
\Delta W=\int_{0}^{\tau}dW=\int_{0}^{\tau}\frac{\partial\phi(x(t),\lambda_{0}(t))}{\partial\lambda_{0}}\frac{d\lambda_{0}(t)}{dt}dt+\int_{0}^{\tau}f(x(t),\lambda_{1}(t))\circ dx
\label{work}
\end{equation}
 and
\begin{equation}
\Delta Q=\int_{0}^{\tau}dQ=\int_{0}^{\tau}\frac{\partial\phi(x(t),\lambda_{0}(t))}{\partial x}\circ dx-\int_{0}^{\tau}f(x(t),\lambda_{1}(t))\circ dx
\end{equation}
 Let us now verify what we expect; that the ratio of conditional path probability densities we use in equation (\ref{ratio})
will be equal to the negative heat transferred to the system divided
by the temperature of the environment. We no longer
have a means for representing the transition probabilities in general
and so we proceed using a so-called `short time propagator' \cite{Risken89,Gardiner09,shorttimepropagator}, which
to first order in the time between transitions, $dt$, describes the
probability of making a transition from $x_{i}$ to $x_{i+1}$. We
may then consider the analysis valid in the limit $dt\to0$. The short
time propagator can also be thought of as a short time Green's function;
it is a solution to the Fokker-Planck equation subject to a delta
function initial condition, valid as the propagation time is taken
to zero. The basic form of the short time propagator is helpfully
rather intuitive and most simply adopts a general gaussian form which
reflects the fluctuating component of the force about the mean due
to the gaussian white noise. Abbreviating ${\cal F}(x,\lambda_{0}(t),\lambda_{1}(t))$
as ${\cal F}(x,t)$ we may write the propagator as
\begin{align}
 & P[x_{i+1},t_{i}+dt|x_{i},t_{i}]=\nonumber \\
 & \sqrt{\frac{m\gamma}{4\pi k_{B}Tdt}}\exp{\left[-\frac{m\gamma}{4k_{B}Tdt}\left(x_{i+1}-x_{i}-\frac{{\cal F}(x_{i},t_{i})}{m\gamma}dt\right)^{2}\right]}.
\label{prop1}
\end{align}
 However, one must be very careful. For reasons similar to those
discussed above, a propagator of this type is not uniquely defined, with a family of forms being available depending on the spatial position at which one chooses to evaluate the force, $\mathcal{F}$, of which equation (\ref{prop1}) is but one example \cite{shorttimepropagator}. In the same way we had to choose certain multiplication rules it is not enough to write $\mathcal{F}(x(t),t)$ on its own since $x(t)$ hasn't been fully specified. This leaves a certain mathematical freedom in how to write the propagator and we must consider which is most appropriate.  Of crucial importance is that all are correct in the limit $dt\to 0$ (all lead to the correct solution of the Fokker-Planck equation) meaning our choice must rest solely on ensuring the correct representation of the entropy production. We can proceed heuristically: as we take time $dt\to 0$ we steadily approach a representation of transitions as jump processes, from which we can proceed with confidence since jump processes are the more general description of stochastic phenomena. In this limit, therefore, we are obliged to faithfully represent the ratio that appears in equation (\ref{ratio}). In that description the forward and reverse jump probabilities
have the same functional form and to emulate this we must evaluate
the short time propagators at the same position $x$ for both the
forward and reverse transition.
\footnote{For the reader aware of the subtleties of stochastic calculus we mention
that for additive noise as considered here this point is made largely
for completeness: if one constructs the result using the relevant stochastic calculus the ratio is independent of the choice. However to be a well defined quantity for cases
involving multiplicative noise this issue becomes important.
}. Mathematically the most convenient way of doing this is to evaluate
all functions in the propagator midway between initial and final points.
Evaluating the functions at the mid-point $x'$ such that $2x'=x_{i+1}+x_{i}$
and $dx=x_{i+1}-x_{i}$ introduces a propagator of the form
\begin{align}
 & P[x_{i+1},t_{i}+dt|x_{i},t_{i}]=\nonumber \\
 & \sqrt{\frac{m\gamma}{4\pi k_{B}Tdt}}\exp{\left[-\frac{m\gamma}{4k_{B}Tdt}\left(\! dx-\frac{{\cal F}(x',t_{i})}{m\gamma}dt\right)^{2}\!-\!\frac{1}{2}\frac{\partial}{\partial x'}\left(\!\frac{{\cal F}(x',t_{i})}{m\gamma}\!\right)dt\right]}
\end{align}
 and similarly
\begin{align}
 & P[x_{i},t_{i}+dt|x_{i+1},t_{i}]=\nonumber \\
 & \sqrt{\frac{m\gamma}{4\pi k_{B}Tdt}}\exp{\left[-\frac{m\gamma}{4k_{B}Tdt}\left(\!-dx-\frac{{\cal F}(x',t_{i})}{m\gamma}dt\right)^{2}\!-\!\frac{1}{2}\frac{\partial}{\partial x'}\left(\!\frac{{\cal F}(x',t_{i})}{m\gamma}\!\right)dt\right]}
\end{align}
 The logarithm of their ratio, in the limit $dt\to0$, simply reduces to
\begin{align}
\lim_{dt\to0}\ln{\left[\frac{P[x_{i+1},t_{i}+dt|x_{i},t_{i}]}{P[x_{i},t_{i}+dt|x_{i+1},t_{i}]}\right]} & =\ln{\left(\frac{P(x_{i+1}|x_{i},\lambda(t_{i}))}{P(x_{i}|x_{i+1},\lambda(t_{i}))}\right)}\nonumber \\
 & =\frac{{\cal F}(x',\lambda_{0}(t_{i}),\lambda_{1}(t_{i}))}{k_{B}T}dx\nonumber \\
 & =\frac{{\cal F}(x(t_{i}),\lambda_{0}(t_{i}),\lambda_{1}(t_{i}))}{k_{B}T}\circ dx\nonumber \\
 & =-\frac{dQ}{k_{B}T}
\end{align}
 where we get to the result by recognising that line two obeys our definition of Stratonovich multiplication rules since $x'$ is the midpoint of $dx$ and that line 3 contains the definition of an increment in the heat transfer from equation (\ref{smallheat}).
We can then construct the entropy production of the entire path by constructing the integral limit of the summation over contributions for each $t_{i}$ such
that
\begin{equation}
k_{B}\ln{\left[\frac{P^{F}[X(\tau)|X(0)]}{P^{R}[\bar{X}(\tau)|\bar{X}(0)]}\right]}=-\frac{1}{T}\int_{0}^{\tau}{dQ}=-\frac{\Delta Q}{T}=\frac{\Delta Q_{\rm med}}{T}=\Delta S_{{\rm med}},
\label{ratio1}
\end{equation}
 giving us the expected result noting that the identification of such a term from the ratio of path probabilities can readily be achieved in full phase space as well \cite{heatfullphase}.

\section{Entropy, Stationarity and Detailed Balance}
\label{detailedbalance:sec}
Let us consider the functional for the total entropy production once
more, specifically with a view to understanding when we expect an
entropy change. Specifically we aim to identify two conceptually different situations where entropy production occurs. If we consider a system evolving without external
driving, it will typically, for well defined system parameters, approach
some stationary state. That stationary state is characterised by a
time independent probability density, $P^{st}$, such that
\begin{equation}
\frac{\partial P^{st}(x,t)}{\partial t}=0.
\end{equation}
 Let us write down the entropy production for such a situation. Since
the system is stationary we have $P_{{\rm start}}=P_{{\rm end}}$,
but we also have a time-independent protocol meaning we need not consider
distinct forward and reverse processes such that we write path probability densities $P^R=P^F=P$. In this situation the total
entropy  production for overdamped motion is given as
\begin{equation}
\Delta S_{{\rm tot}}[x]=k_{B}\ln{\left[\frac{P^{st}(x(0))P[x(\tau)|x(0)]}{P^{st}(x(\tau))P[x(0)|x(\tau)]}\right]}.
\end{equation}
 We can then
ask what in general are the properties required for entropy production,
or indeed no entropy production in such a situation. Clearly there
is no entropy production when the forward and reverse trajectories
are equally likely and so we can write the condition for zero entropy
production in the stationary state as
\begin{equation}
P^{st}(x(0))P[x(\tau)|x(0)]=P^{st}(x(\tau))P[x(0)|x(\tau)]\qquad\forall\; x(0),x(\tau).
\label{detailedbalance}
\end{equation}
 Written in this form we emphasise that this is equivalent to the
statement of \emph{detailed balance}. Transitions are said to balance because
the average number of all transitions to and from any given configuration
$x(0)$ exactly cancel; this leads to a constant probability distribution
and is the condition required for a stationary state. However to have
no entropy production in the stationary state we require all transitions
to balance in detail: we require the total number of transitions between
every possible combination of two configurations $x(0)$ and $x(\tau)$
to cancel. This is also the condition required for zero probability current and for the system to be at thermal equilibrium where we understand the entropy
of the universe to be maximised.\\
\\
We may then quite generally place any dynamical scheme into one of two broad categories. The first is where detailed balance (equation (\ref{detailedbalance})) holds and the stationary state is the thermal equilibrium \footnote{ One can build models which have stationary states that have zero entropy production where equilibrium is only local, but there is no value in distinguishing between the two situations or highlighting such cases here.}. Under such dynamics systems left unperturbed will relax towards equilibrium where there is no observed preferential forward or reverse behaviour, no observed thermodynamic arrow of time or irreversibility and therefore no entropy production. Thus all entropy production for these dynamics is the result of driving and subsequent relaxation to equilibrium or more generally as a consequence of the systems being out of their stationary states.\\
\\
The other category therefore is where detailed balance does not hold. In these situations we expect entropy production even in the stationary state which by extension must have origins beyond that of driving out of and relaxation back to stationarity. So when can we expect detailed balance to be broken? We can first identify the situations where it does hold and for overdamped motion, the requirements are well defined. To have all transitions balancing
in detail is to have zero probability current, $J^{{\rm st}}(x,t)=0$, in the stationary
state, where the current is related to the probability density according to
\begin{equation}
\frac{\partial P^{st}(x,t)}{\partial t}=-\frac{\partial J^{st}(x,t)}{\partial x}=0.
\end{equation}
 Utilising the form of the Fokker-Planck equation that corresponds
to the dynamics we would thus require
\begin{equation}
J^{st}(x,t)=\frac{1}{m\gamma}\left(-\frac{\partial\phi(x,\lambda_{0}(t))}{\partial x}+f(x,\lambda_{1}(t))\right)P^{st}(x,t)-\frac{k_{B}T}{m\gamma}\frac{\partial P^{st}(x,t)}{\partial x}=0.
\end{equation}
 We can verify the consistency of such a condition by inserting the
appropriate stationary distribution
\begin{equation}
P^{st}(x,t)\propto\exp{\left[\int^{x}dx'\;\frac{m\gamma}{k_{B}T}\left(-\frac{\partial\phi(x',\lambda_{0}(t))}{\partial x'}+f(x',\lambda_{1}(t))\right)\right]}
\label{stationary}
\end{equation}
 which is clearly of a canonical form. How might one break this condition? We would require a non-vanishing current
and this can be achieved when the contents of the exponential in equation
(\ref{stationary}) are not integrable. In general this can be achieved
by using an external force that is non-conservative. However in one
dimension with natural, that is reflecting, boundary conditions any
force acts conservatively since the total distance between initial and final positions, and thus work done is always path
independent. To enable such a non-conservative force one can implement
periodic boundary conditions. This might be realised physically by
considering motion on a ring since when a constant force acts on the particle
the work done will depend on the number of times the particle traverses
that ring. If the system relaxes to its stationary state there will
be a non-zero, but constant current that arises due to the non-conservative
force driving the motion in one direction. In such
a system with steady flow it is quite easy to understand that the
transitions between two configurations will not cancel and
thus detailed balance is not achieved. Allowing these dynamics to relax the system to its stationary state creates a simple
example of a \emph{non-equilibrium steady state}. Generally such states can be created by the placing of some constraint upon the system
which stops it from reaching a thermal equilibrium. This results in a system  which
is perpetually attempting  and failing to maximise the total entropy by equilibrating. By remaining out of equilibrium it constantly dissipates heat to the environment and is thus associated with a constant
entropy generation. As such, a system with these dynamics gives rise to  irreversibility beyond that arising from driving and relaxation and possesses an underlying breakage of time reversal symmetry, leading to an associated entropy production, manifest in the lack of detailed balance. Detailed balance may be broken in many ways and the non-equilibrium constraint that causes it may be, as we have seen, a non-conservative
force, or it might be exposure to particle reservoirs with unequal
chemical potentials or heat baths with unequal temperatures. The steady states of such systems in particular are of great interest in statistical physics, not only because of their qualitatively different behaviour, but also because they provide cases where analytical solution is feasible out of equilibrium. As we shall see later the distribution of entropy production in  these states also obeys a particular powerful symmetry requirement.

\section{A general fluctuation theorem}

So far we have examined a particular functional of
a path and argued from a number of perspectives that it represents the
total entropy production of the universe. We have also seen that it
obeys a remarkably simple and powerful relation which guarantees its
positivity on average. However, we can exploit the form of the entropy
production further and derive a number of fluctuation theorems which explicitly
relate distributions of general entropy-like quantities. They are
numerous and the differences can appear rather subtle, however it is
quite simple to derive a very general equality which we can rigorously
and systematically adapt to different situations and arrive at these
different relations. To do so let us once again consider the functional
which represents the total entropy production
\begin{equation}
\Delta S_{{\rm tot}}[X]=k_{B}\ln{\left[\frac{P_{{\rm start}}(X(0))P^{F}[X(\tau)|X(0)]}{P_{{\rm start}}^{R}(\bar{X}(0))P^{R}[\bar{X}(\tau)|\bar{X}(0)]}\right]}.
\end{equation}
 We are able to construct the probability distribution of this quantity
 for a particular process. Mathematically, the distribution
of entropy production over the forward process can be written
as
\begin{equation}
P^{F}(\Delta S_{{\rm tot}}[X]=A)=\int dX\; P_{{\rm start}}(X(0))P^{F}[X(\tau)|X(0)]\delta(A-\Delta S_{{\rm tot}}[X])
\end{equation}
 To proceed we follow Harris et al. \cite{Harris07} and consider a new functional,
but one which is very similar to the total entropy production. We
shall generally refer to it as $R$ and it can be written
\begin{equation}
R[X]=k_{B}\ln{\left[\frac{P_{{\rm start}}^{R}(X(0))P^{R}[X(\tau)|X(0)]}{P_{{\rm start}}(\bar{X}(0))P^{F}[\bar{X}(\tau)|\bar{X}(0)]}\right]}.
\end{equation}
 Imagine that we evaluate this new quantity over the reverse trajectory,
that is we consider $R[\bar{X}]$. It will be given by
\begin{equation}
R[\bar{X}]=k_{B}\ln{\left[\frac{P_{{\rm start}}^{R}(\bar{X}(0))P^{R}[\bar{X}(\tau)|\bar{X}(0)]}{P_{{\rm start}}({X}(0))P^{F}[{X}(\tau)|{X}(0)]}\right]}=-\Delta S_{{\rm tot}}[X]
\end{equation}
 which is explicitly the negative value of the functional that represents
the total entropy production in the forward process. We can similarly
construct a distribution for $R[\bar{X}]$ over the reverse process.
This in turn would be given as
\begin{equation}
P^{R}(R[\bar{X}]=A)=\int d\bar{X}\; P_{{\rm start}}^{R}(\bar{X}(0))P^{R}[\bar{X}(\tau)|\bar{X}(0)]\delta(A-R[\bar{X}]).
\end{equation}
 We now seek to relate this distribution to that of the total
entropy production over the forward process. To do so we consider
the value the probability distribution takes for $R[\bar{X}]=-A$.
By the symmetry of the delta function we may write
\begin{equation}
P^{R}(R[\bar{X}]=-A)=\int d\bar{X}\; P_{{\rm start}}^{R}(\bar{X}(0))P^{R}[\bar{X}(\tau)|\bar{X}(0)]\delta(A+R[\bar{X}]).
\end{equation}
 We now utilise three substitutions. First, $dX=d\bar{X}$ denoting
the equivalence of measure as the trajectories are defined on the same
space. Next we use the definition of the entropy production functional
to substitute
\begin{equation}
P_{{\rm start}}^{R}(\bar{X}(0))P^{R}[\bar{X}(\tau)|\bar{X}(0)]=P_{{\rm start}}(X(0))P^{F}[X(\tau)|X(0)]{\rm e}^{-\Delta S_{{\rm tot}}[X]/k_{B}}
\end{equation}
 and finally the definition that $R[\bar{X}]=-\Delta S_{{\rm tot}}[X]$.
Performing the above substitutions we find
\begin{align}
\!\!\!P^{R}(R[\bar{X}]\!=\!-A) & \!=\!\!\int\!\! d{X}\, P_{{\rm start}}({X}(0))P^{F}[{X}(\tau)|{X}(0)]{\rm e}^{-\frac{\Delta S_{{\rm tot}}[X]}{k_{B}}}\delta(A\!-\!\Delta S_{{\rm tot}}[{X}])\nonumber \\
 & \!={\rm e}^{-\frac{A}{k_{B}}}\int\! d{X}\, P_{{\rm start}}({X}(0))P^{F}[{X}(\tau)|{X}(0)]\delta(A\!-\!\Delta S_{{\rm tot}}[{X}])\nonumber \\
 & \!={\rm e}^{-\frac{A}{k_{B}}}\,P^{F}(\Delta S_{{\rm tot}}[{X}]\!=\! A)
\end{align}
 and yields \cite{Harris07}\\
 \\
 { \emph{The Transient Fluctuation Theorem}}:
\begin{equation}
P^{R}(R[\bar{X}]=-A)={\rm e}^{-\frac{A}{k_{B}}}\,P^{F}(\Delta S_{{\rm tot}}[{X}]=A)
\label{transient}
\end{equation}
 This is a fundamental relation and holds for all protocols and initial
conditions and is of a form referred to in the literature as a finite time,
transient or detailed fluctuation theorem depending on where you look.
Additionally, if we integrate over all values of $A$ on both sides
we obtain the integral fluctuation theorem
\begin{equation}
1=\langle{\rm e}^{-\Delta S_{{\rm tot}}/k_{B}}\rangle
\end{equation}
 with its name now being self-explanatory. These two relations shall
now form the basis of all relations we consider. However, upon returning
to the transient fluctuation theorem, a valid question is what does
the functional $R[\bar{X}]$ represent? In terms of traditional thermodynamic
quantities there is scant physical interpretation. It is more helpful
to consider it as a related functional of the path and to understand
that in general it is \emph{not} the entropy production of the
reverse path in the reverse process. It is important now to look at
why. To construct the entropy production under the reverse process,
we need to consider a new functional which we shall call $\Delta S_{{\rm tot}}^{R}[\bar{X}]$
that is defined in exactly the same way as for the forward process.
We consider an initial distribution, this time $P_{{\rm start}}^{R}$
which evolves to $P_{{\rm end}}^{R}$ and compare the probability density for a trajectory starting from the initial distribution, this time
under the reverse protocol $\bar{\lambda}(t)$, with the probability density of a trajectory starting from the time reversed final distribution,
$\hat{T}P_{{\rm end}}^{R}$, so that
\begin{equation}
\Delta S_{{\rm tot}}^{R}[\bar{X}]=k_{B}\ln{\left[\frac{P_{{\rm start}}^{R}(\bar{X}(0))P^{R}[\bar{X}(\tau)|\bar{X}(0)]}{\hat{T}P_{{\rm end}}^{R}({X}(0))P^{F}[{X}(\tau)|{X}(0)]}\right]}\neq-\Delta S_{{\rm tot}}[X].\label{eq:83a}
\end{equation}
 Crucially there is an inequality in equation (\ref{eq:83a}) in general
because
\begin{equation}
\hat{T}P_{{\rm start}}(X(0)))\neq P_{{\rm end}}^{R}(\bar{X}(\tau))=\int d\bar{X}\; P_{{\rm start}}^{R}(\bar{X}(0))P^{R}[\bar{X}(\tau)|\bar{X}(0)].\label{eq:83}
\end{equation}
 This is manifest in the irreversibility of the dynamics of the systems
we are looking at, as is illustrated in Figure \ref{forback}. If
the dynamics were reversible, as for Hamilton's equations and Liouville's
theorem, then equation (\ref{eq:83}) would hold in equality. So,
examining equations (\ref{transient}) and (\ref{eq:83a}), if we wish to compare the
distribution of entropy production in the reverse process with that for the forward
process, we need to have $R[\bar{X}]=\Delta S_{{\rm tot}}^{R}[\bar{X}]$
such that $\Delta S_{{\rm tot}}[X]=-\Delta S_{{\rm tot}}^{R}[\bar{X}]$.
This is achieved by having $P_{{\rm start}}({X}(0))=\hat{T}P_{{\rm end}}^{R}(\bar{X}(\tau))$.
When this condition is met we may write
\begin{equation}
P^{R}(\Delta S_{{\rm tot}}^{R}[\bar{X}]=-A)={\rm e}^{-\frac{A}{k_{B}}}P^{F}(\Delta S_{{\rm tot}}[{X}]=A)
\end{equation}
 which now relates distributions of the same physical quantity, entropy
change. If we assume that arguments of a probability distribution for the reverse protocol
$P^{R}$ implicitly describe the quantity over the reverse
process we may write it in its more common form
\begin{equation}
P^{R}(-\Delta S_{{\rm tot}})={\rm e}^{-\Delta S_{{\rm tot}}/k_{B}}P^{F}(\Delta S_{{\rm tot}})
\label{DFTSeif}
\end{equation}
 This will hold when the protocol and initial distributions are chosen
such that evolution under the forward process followed by the reverse
process together with the appropriate time reversals brings the system back
into the same initial statistical distribution. This sounds somewhat
challenging and indeed does not occur in any generality, but there
are two particularly pertinent situations where the above does hold
and has particular relevance in a discussion of thermodynamic quantities.

\subsection{Work Relations}
\label{workrelations:sec}
The first and most readily applicable example that obeys the condition
$P_{{\rm start}}({X}(0))=\hat{T}P_{{\rm end}}^{R}(\bar{X}(\tau))$
is that of changes between equilibrium states where one can trivially
obtain the required condition by exploiting the fact that unperturbed,
the dynamics will steadily bring the system into a stationary state
which is invariant under time reversal. We start by defining the equilibrium
distribution which represents the canonical ensemble where, as before,
we consider the system energy for an overdamped system to be entirely
described by the potential $\phi(x,\lambda_{0}(t))$ such that
\begin{equation}
P^{eq}(x(t),\lambda_{0}(t))=\frac{1}{Z({\lambda_{0}(t)})}\exp\left[-\frac{\phi(x(t),\lambda_{0}(t))}{k_{B}T}\right].
\end{equation}
for $t=0$ and $\tau$, where $Z$ is the partition function, uniquely
defined by $\lambda_{0}(t)$, which can in general be related to the
Helmholtz free energy through the relation
\begin{equation}
F(\lambda_{0}(t))=-k_{B}T\ln{Z(\lambda_{0}(t))}.
\end{equation}
 To clarify, the corollary of these statements is to say that the
directly applied force $f(x(t),\lambda_{1}(t))$ does not feature
in the system's Hamiltonian%
\footnote{That is not to say it may not appear in some generalised Hamiltonian.
For further insight into this issue we direct the interested reader
to, for example  \cite{Jarworkrelations,Jarworkrelations0}, noting the approach here
and elsewhere  \cite{seifertprinciples} best resembles the extended relation used in  \cite{Jarworkrelations}.
}. Let us now choose the initial and final distributions to be given
by the respective equilibria defined by the protocol at the start
and finish of the forward process and the same temperature
\begin{align}
P_{{\rm start}}(x(0),\lambda_{0}(0)) & \propto\exp\left[\frac{F(\lambda_{0}(0))-\phi(x(0),\lambda_{0}(0))}{k_{B}T}\right]\nonumber \\
P_{{\rm end}}(x(\tau),\lambda_{0}(\tau)) & \propto\exp\left[\frac{F(\lambda_{0}(\tau))-\phi(x(\tau),\lambda_{0}(\tau))}{k_{B}T}\right].
\end{align}
 We are now in a position to construct the total entropy change for
a given realisation of the dynamics between these two states. From
the initial and final distributions we can immediately construct the
system entropy change $\Delta S_{{\rm sys}}$ as
\begin{align}
\Delta S_{{\rm sys}} & =k_{B}\ln{\left(\frac{P_{{\rm start}}(x(0),\lambda_{0}(0))}{P_{{\rm end}}(x(\tau),\lambda_{0}(\tau))}\right)}=k_{B}\ln{\left(\frac{\exp\left[\frac{F(\lambda_{0}(0))-\phi(x(0),\lambda_{0}(0))}{k_{B}T}\right]}{\exp\left[\frac{F(\lambda_{0}(\tau))-\phi(x(\tau),\lambda_{0}(\tau))}{k_{B}T}\right]}\right)}\nonumber \\
 & =\frac{1}{T}\left(-F(\lambda_{0}(\tau))+F(\lambda_{0}(0))+\phi(x(\tau),\lambda_{0}(\tau))-\phi(x(0),\lambda_{0}(0))\right)\nonumber \\
 & =\frac{\Delta\phi-\Delta F}{T}
\end{align}
 The medium entropy change is as we defined previously and can be
written
\begin{equation}
\Delta S_{{\rm med}}=-\frac{\Delta Q}{T}=\frac{\Delta W-\Delta\phi}{T}
\end{equation}
 where $\Delta W$ is the work given earlier in equation (\ref{work}), but we now emphasise
that this term contains contributions due to changes in the potential
and due to the external force $f$. We thus further define two new quantities
$\Delta W_{0}$ and $\Delta W_{1}$ such that $\Delta W=\Delta W_{0}+\Delta W_{1}$
with
\begin{equation}
\Delta W_{0}=\int_{0}^{\tau}\frac{\partial\phi(x(t),\lambda_{0}(t))}{\partial\lambda_{0}}\frac{d\lambda_{0}(t)}{dt}dt
\label{W0}
\end{equation}
 and
\begin{equation}
\Delta W_{1}=\int_{0}^{\tau}f(x(t),\lambda_{1}(t))\circ dx.
\label{W1}
\end{equation}
$W_0$ and $W_1$ are not defined in the same way with $W_0$ being found more often in thermodynamics and $W_1$ being a familiar definition from mechanics: One may therefore refer to these definitions as thermodynamic and mechanical work respectively. The total entropy production in this case is simply given by
\begin{equation}
\Delta S_{{\rm tot}}[x]=\frac{\Delta W-\Delta F}{T}.
\end{equation}
 Additionally, since we have established that $P_{{\rm end}}^{R}(\bar{x}(\tau))=\hat{T}P_{{\rm start}}(x(0))$
we can also write
\begin{equation}
\Delta S_{{\rm tot}}^{R}[\bar{x}]=-\frac{\Delta W-\Delta F}{T}.
\end{equation}

\subsubsection{The Crooks Work Relation and Jarzynski Equality}

\begin{figure}[!htp]
\begin{centering}
\includegraphics[clip,width=100mm]{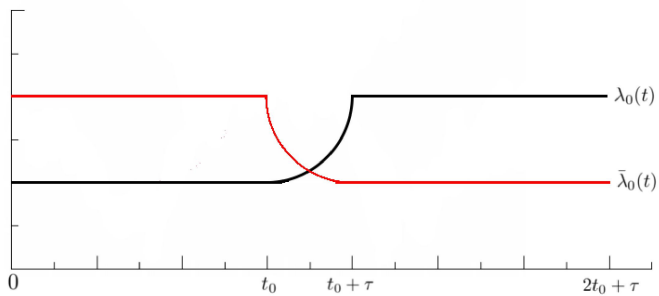}
 \caption{A protocol $\lambda_{0}(t)$, of duration $2t_{0}+\tau$, which evolves
a system from one equilibrium to another defined so that the reversed
protocol $\bar{\lambda}_{0}(t)$ returns the system to the original
equilibrium. There is a period of no driving of length $t_{0}$ which
corresponds to the relaxation for the reverse process, followed by
a time $\tau$ of driving, followed by another relaxation period of
duration $t_{0}$. As we take $t_{0}$ to infinity we obtain a protocol
which produces the condition $P_{{\rm end}}^{R}(\bar{x}(2t_{0}+\tau))=\hat{T}P_{{\rm start}}(x(0))$
. We note here that $f(x(t),\lambda_{1}(t))=0$.}
\label{jarfig}
\end{centering}
\end{figure}

The derivation of several relations follows now by imposing certain
constraints on the process we consider. First let us imagine the situation
where the external force $f(x,\lambda_{1})=0$ and so all work is
performed conservatively through the potential such that $\Delta W=\Delta W_{0}$.
To proceed we should clarify the form of the protocol that would take
an equilibrium system to a new equilibrium such that its reversed
counterpart would return it to the same initial distribution. This
would consist of a waiting period, in principle of infinite duration,
where the protocol is constant, followed by a period of driving where
the protocol changes, followed by another infinitely long waiting
period. Such a protocol is given and explained in Figure \ref{jarfig}.

For such a process we write the total entropy production
\begin{equation}
\Delta S_{{\rm tot}}=\frac{\Delta W_0-\Delta F}{T}.
\end{equation}
 This changes its sign for the reverse trajectory and reverse protocol
and so we may construct the appropriate fluctuation relation which
is now simply read off equation (\ref{transient}) as
\begin{equation}
P^{F}\left(({\Delta W_0-\Delta F})/{T}\right)=\exp{\left[\frac{\Delta W_0-\Delta F}{k_{B}T}\right]}P^{R}\left(-({\Delta W_0-\Delta F})/{T}\right).
\end{equation}
 Since $F$ and $T$ are independent of the trajectory we can simplify
and find \cite{Crooks99}\\
 \\
 \emph{The Crooks Work Relation}:
\begin{equation}
\frac{P^{F}\left(\Delta W_0\right)}{P^{R}\left(-\Delta W_0\right)}=\exp{\left[\frac{\Delta W_0-\Delta F}{k_{B}T}\right]}.
\end{equation}
 \\
Rearranging and integrating over all $\Delta W$ on both sides, and
taking the deterministic $\Delta F$ out of the path integral then
yields an expression for the average over the forward process called \cite{Jarzynski97,jaroriginal1,Crooks98}\\
 \\
 \emph{The Jarzynski Equality}:
\begin{equation}
\left\langle \exp{\left({-\Delta W_0}/{k_{B}T}\right)}\right\rangle =\exp{\left(-\Delta F/k_{B}T\right)}
\label{Jareq}
\end{equation}
 \\
 The power of these statements is clarified in one very important
conceptual point. In their formulation the relations are \emph{constructed}
using the values of entropy change for a process which, after starting
in equilibrium, is isolated for a long time, driven and then left
for a long time again to return to a stationary state. However this
does not mean that these quantities have to be \emph{measured} over
the whole of such a process. Why is this the case? It is because the
entropy production for the whole process can be written in terms of
the mechanical work and free energy change which are delivered exclusively
during the driving phase when the protocol $\lambda_{0}(t)$ is changing.
Since the work and free energy change are independent of the intervals
where the protocol is constant and because we had no constraint on
$\lambda_{0}(t)$ during the driving phase we can therefore consider
them to be valid for any protocol assuming the system is in equilibrium
to start with. We can therefore state that the Crooks work relation
and Jarzynski equality hold \emph{for all times} for systems that
start in equilibrium%
\footnote{Although we have only shown that this is the case for Langevin dynamics,
it is important to note that these expressions can be obtained for
other general descriptions of the dynamics. See for example \cite{Jarstrong}.%
}. Historically this has had one particularly important consequence: the
results hold for driving, in principle, arbitrarily far from equilibrium.
This is widely summed up as the ability to obtain equilibrium information
from non-equilibrium averaging since, upon examining the form of the
Jarzynski equality, we can compute the free energy difference by taking
an average of the exponentiated work done in the course of some non-equilibrium
process. Exploiting these facts let us clarify what these two relations
mean explicitly and what the implications are in the real world.
\\\\
\emph{The Crooks Relation}
\begin{quotation}
 \noindent Statement:\\
  For any time $\tau$, the probability of observing trajectories which
correspond to an application of $\Delta W_0$ work, starting from an
equilibrium state defined by $\lambda(0)$, under dynamics described
by $\lambda(t)$ in $0\le t\le\tau$, is exponentially more likely
in $(\Delta W_0-\Delta F)/k_{B}T$ than the probability of observing
trajectories that correspond to an application of $-\Delta W_0$ work
from an equilibrium state defined by $\bar{\lambda}(0)$, under dynamics
described by $\bar{\lambda}(t)$.
\\\\
 Implication:\\
 Consider an isothermal gas in a piston in contact with a heat bath
at equilibrium. Classically we know from the second law that if we
compress the gas, performing a given amount of work $\Delta W_0$ on
it, then after an equilibration period, we must expect the gas to
perform an amount of work that is less than $\Delta W_0$ when it is
expanded (i.e. $-\Delta W_0$ work performed on the gas). To get the
same amount of work back out we need to perform the process quasistatically
such that it is reversible. The Crooks relation however, tells us
more. For the same example, we can state that if the dynamics of our
system lead to some probability of performing $\Delta W_0$ work, then
the probability of extracting the same amount of work in the reverse
process differs exponentially. Indeed they only have the same probability
when the work performed is equal to the free energy difference, often
called the reversible work.
\end{quotation}
\emph{The Jarzynski Equality}
\begin{quotation}
\noindent Statement:\\
 For any time $\tau$ the average value, as defined by the mean over
many realisations of the dynamics, of the exponential of the negative
work divided by the temperature arising from a defined change in protocol
from $\lambda_{0}(0)$ to $\lambda_{0}(\tau)$ is identically equal
to the exponential of the negative equilibrium free energy difference corresponding
to the same change in protocol, divided by the temperature.\\\\
Implication:\\
Consider once again the compression of a gas in a piston, but let us imagine that we
wish to know the free energy change without knowledge of the equation
of state. Classically, we might be able to measure the free energy
change by attempting to perform the compression quasistatically; which
of course can never be fully realised. However, the Jarzynski equality
states that we can determine this free energy change exactly by repeatedly compressing
the gas \emph{at any speed} and taking an average of the exponentiated
work that we perform over all these fast compressions. One must exercise
caution however; the average taken is patently dominated by very negative
values of work. These correspond to very negative excursions in entropy
and are often \emph{extremely} rare. One may find that the estimated
free energy change is significantly altered following one additional
realisation even if hundreds or perhaps thousands have already been
averaged.
\end{quotation}

\noindent These relations very concisely extend the classical definition of
irreversibility in such isothermal systems. In classical thermodynamics
we may identify the difference in free energy as the maximum amount
of work we may extract from the system, or rather that to achieve
a given free energy change we must perform at least as much work as
that free energy change, that is
\begin{equation}
\Delta W_0\geq\Delta F
\end{equation}
 with the equality holding for a quasistatic `reversible' process.
But just as we saw that our entropy functional could take negative
values there is nothing in the dynamics which prevents an outcome
where the work is less than the free energy change. We understand
now that the second law is statistical so more generally we must have
\begin{equation}
\langle\Delta W_0\rangle\geq\Delta F.
\end{equation}
 The Jarzynski equality tells us more than this and replaces the inequality
with an equality that it is valid for non-quasistatic processes where
mechanical work is performed at a finite rate such that the system
is driven away from thermal equilibrium and the process is irreversible.

\subsection{Fluctuation relations for mechanical work}

\begin{figure}[!htp]
\begin{centering}
\includegraphics[clip,width=100mm]{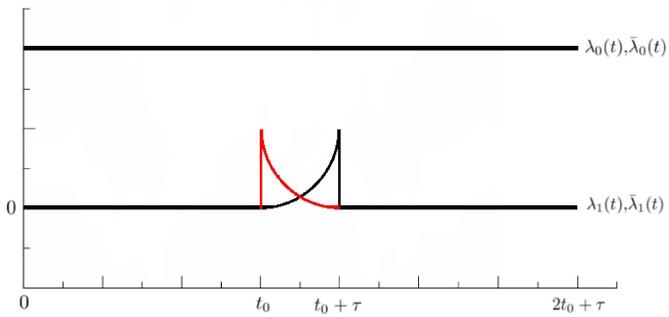}
\caption{An example protocol and reversed protocol that would construct the
condition $P_{{\rm end}}^{R}(\bar{x}(2t_{0}+\tau))=\hat{T}P_{{\rm start}}(x(0))$
when all work is performed through the external force $f(x(t),\lambda_{1}(t))=\lambda_{1}(t)$
and $t_{0}$ is taken to infinity.}
\label{bochfig}
\end{centering}
\end{figure}

Let us now consider a similar, but subtly different circumstance to
that of the Jarzynski and Crooks relations. We consider a driving
process that again starts in equilibrium, but this time keeps the
protocol $\lambda_{0}(t)$ held fixed such that all work is performed
by the externally applied force $f(x(t),\lambda_{1}(t))$ meaning
that $\Delta W=\Delta W_{1}$. Once again we seek a fluctuation relation
by constructing an equilibrium to equilibrium process, though this
time we insist that the system relaxes back to the same initial equilibrium
distribution. We note that since $f(x(t),\lambda_{1}(t))$ may act
non-conservatively, in order to allow relaxation back to equilibrium
we would require that the external force be `turned off'. An example
set of protocols is given in Figure \ref{bochfig} for a simple external
force $f(x(t),\lambda_{1}(t))=\lambda_{1}(t)$.

For such a process we find
\begin{equation}
\Delta S_{{\rm tot}}=\frac{\Delta W_{1}}{T}
\end{equation}
 since the free energy difference between the same equilibrium states
vanishes. We have constructed a process such that the distribution at the end
of the reverse process is (with time reversal) the same as the initial distribution of
the forward process and so again we are permitted to read off a set
of fluctuation relations \cite{Jarworkrelations,Jarworkrelations0,Bochkov81,Bochkov2} which may collectively be referred to as\\
 \\
 \emph{Fluctuation relations for mechanical work:}
\begin{equation}
\frac{{\cal P}^{F}(\Delta W_{1})}{{\cal P}^{R}(-\Delta W_{1})}=\exp{\left[\frac{\Delta W_{1}}{k_{B}T}\right]}
\label{FRmech}
\end{equation}

\begin{equation}
\langle\exp{(-\Delta W_{1}/k_{B}T)}\rangle=1
\end{equation}
 \\
 For the same reasons as in the Jarzynski and Crooks relations they
are valid for all times and thus hold as a non-equilibrium result.
Taking in particular the integrated relation and
comparing with the Jarzynski equality in equation (\ref{Jareq}) one may think there is
an inconsistency. Both are valid for all times and arbitrary driving
and concern the work done under the constraint that both start
in equilibrium, yet on first inspection they seem to be saying different
things. But recall our distinction between the work $\Delta W_{0}$
and $\Delta W_{1}$ from equations (\ref{W0}) and (\ref{W1}); there are two distinct
ways to describe work on such a particle. If one performs work $\Delta W_{0}$
one necessarily changes the form of the system energy whereas the
application of work $\Delta W_{1}$ leaves the form of the system
energy unchanged. The difference is manifest in the two different
integrated relations because their derivations exploit the fact that
the Hamiltonian, which represents the system energy, appears in initial
and final distributions. To clarify, as written the Jarzynski equality
explicitly concerns driving where the application of any work also
changes the Hamiltonian and thus the equilibrium state. On the other
hand the relations for $W_1$ concern work as the path integral
of an external force such that the Hamiltonian remains unchanged for
the entire process.\\
 \\
 Of course, there is nothing in the derivation of either of these
relations that precludes the possibility of both types of work to
be performed at the same time and so using the same arguments we arrive at
\begin{equation}
\frac{P^{F}(\Delta W)}{P^{R}(-\Delta W)}=\exp{\left[(\Delta W-\Delta F)/k_{B}T\right]}
\end{equation}
 and
\begin{equation}
\langle{\rm e}^{-\frac{\Delta W-\Delta F}{k_{B}T}}\rangle=1
\end{equation}
 again under the constraint that the system be initially
prepared in equilibrium.

\subsection{Fluctuation theorems for entropy production}
\label{FTentropy:sec}
We have seen in section \ref{workrelations:sec} how relations between distributions of
work can be derived from equation (\ref{DFTSeif}), since work can be related to the entropy production
during a suitable equilibrium to equilibrium process. We wish now
to seek situations where we can explicitly construct relations that
concern the  distributions of the entropy produced for given
forward and reverse processes that do not necessarily begin and end in equilibrium. In order
to find situations where the value of the entropy production for the forward trajectory in the
forward process is precisely the negative value of the entropy production
for the reversed trajectory in the reverse process, we seek situations
where the reverse protocol acts to return the distribution to the
time reversed initial distribution for the forward process. For the overdamped motion we have
been considering we would require
\begin{equation}
P_{{\rm start}}(x(\tau))=\int dx\; P_{{\rm end}}(x(0))P^{R}\left[x(\tau)|x(0)\right].
\end{equation}
  We seek a situation different from the previously considered equilibrium to equilibrium process,  occurring when the  functional forms of the initial and final distributions are the same such that $P_{{\rm start}}=P_{{\rm end}}$.
 One can expect to see a symmetry between distributions of entropy production in such forward and reverse processes along the lines of equation (\ref{DFTSeif}).

However, we can specify further and find an even more direct symmetry if we insist that the evolution under the forward process
is indistinguishable from that under the reverse process. Mathematically this
means $P(x_{i+1}|x_{i},\lambda(t_{i+1}))=P(x_{i+1}|x_{i},\bar{\lambda}(t_{i+1}))$
or $P^{R}[x(\tau)|x(0)]=P^{F}[x(\tau)|x(0)]$. Given these conditions
evolution from the initial distribution will result in the final distribution
and evolution under the reverse process from the final distribution
will result in the initial distribution. If we consider in more detail the requirements for such a condition
we understand there are two main ways in which this can be achieved.
Given that the initial and final distributions are the same the first
way is to require a constant protocol $\lambda(t)$. In this way the
forward process is trivially the same as the reverse process. Alternatively we require
the protocol to be time symmetric such that $\lambda(t)=\lambda(\tau-t)=\bar{\lambda}(t)$.
In both situations the forward and reverse processes
are entirely indistinguishable.
As such, by careful construction we can, in these specific circumstances,
relate the probability of seeing a positive entropy production to
that of a negative entropy production over the \emph{same forward
process} allowing us from equation (\ref{transient}) to write a \cite{seifertoriginal}
\\
\emph{Detailed fluctuation theorem:}\\
\begin{equation}
P(\Delta S_{{\rm tot}})={\rm e}^{\Delta S_{{\rm tot}}/k_{B}}P(-\Delta S_{{\rm tot}})
\label{DFTgen}
\end{equation}
 Physically the two situations we have considered correspond to
\begin{itemize}
\item $P_{{\rm start}}=P_{{\rm end}}$, $\lambda(t)=\text{const}$:\\
 To satisfy such criteria the system must be in a steady state, that
is all intrinsic system properties (probability distribution, mean
system entropy, mean system energy etc) must remain constant over
the process. The simplest steady state is equilibrium which trivially
has zero entropy production in detail for all trajectories. However,
a \emph{non-equilibrium} steady state can be achieved by breaking
detailed balance through some constraint which prevents the equilibration
as we saw in section \ref{detailedbalance:sec}. The mean entropy production rate of these states is constant, non-zero
and, as we have now shown, there is an explicit exponential symmetry in
the probability of positive and negative fluctuations.
\item $P_{{\rm start}}=P_{{\rm end}}=P$, $\lambda(t)=\bar{\lambda}(t)$:\\
 This condition can be achieved in a system that is being periodically
driven characterised by a time symmetric $\lambda(t)$. If from some
starting point we allow the system to undergo an arbitrarily large
number of periods of driving it will arrive at a so-called non-equilibrium
oscillatory state such that $P(x,t)=P(x,t+t_{p})$ where $t_{p}$
is the period of oscillation. In this state we can expect the above
relation to hold for integer multiples of period $t_{p}$ starting
from a time such that $\lambda(t)=\bar{\lambda}(t)$.
\end{itemize}

\section{Further results}

\subsection{Asymptotic fluctuation theorems}

In the class of system we have considered, the integral and detailed
fluctuation theorems are guaranteed to hold. Indeed it has not escaped
some authors' attention that the reason they do is fully explained
in the very definition of the functionals they concern \cite{shargel}. There is, however, a class
of fluctuation theorems which does not have this property. These are
known as asymptotic fluctuation theorems. Their derivation for Langevin and then general Markovian stochastic systems is due to Kurchan  \cite{Kurchan} and Lebowitz and Spohn  \cite{GCforstochastic} respectively, and superficially bear strong similarities with results obtained by Gallavotti and Cohen for chaotic deterministic systems  \cite{Gallavotti95,Gallavotti2}.
They generally concern systems that approach a steady state, and, for stochastic systems, strictly
in their definition concern a symmetry in the long time limit of the
generating function of a quantity known as an action functional or
flux \cite{GCforstochastic}. This quantity again concerns a trajectory that runs from $x_{0}$
through to $x_{n}$, described using jump probabilities $\sigma(x_{i}|x_{i-1})$
and is given as
\begin{equation}
\mathcal{W}(t)=\ln{\left[\frac{\sigma(x_{1}|x_{0})}{\sigma(x_{0}|x_{1})}\ldots\frac{\sigma(x_{n}|x_{n-1})}{\sigma(x_{n-1}|x_{n})}\right]}.
\end{equation}
 The statement of such an asymptotic fluctuation theorem, which
we shall not prove and only briefly address here, states that  there exists a long time limit of the scaled cumulant generating function of $\mathcal{W}(t)$ such that
\begin{equation}
e(s)=\lim_{t\to\infty}-\frac{1}{t}\ln{\langle\exp{\left[-s\mathcal{W}(t)\right]}\rangle}
\end{equation}
 and that this quantity possesses the symmetry
\begin{equation}
e(s)=e(1-s).
\label{GCsym}
\end{equation}
 From this somewhat technical definition we can derive a fluctuation
theorem closely related to those which we have examined already. The
existence of such a limit implies that the distribution function of  the time averaged action functional
$\mathcal{W}(t)/t$, ${P}(\mathcal{W}(t)/t)$, follows a large
deviation behaviour \cite{hugo} such that, in the long time limit we have
\begin{equation}
 {P}(\mathcal{W}(t)/t)\simeq{\rm e}^{-t\hat{e}(\mathcal{W}/t)}
\label{largedev}
\end{equation}
 where $\hat{e}$ is the Legendre transform of $e$ defined as
\begin{equation}
\hat{e}(\mathcal{W}/t)=\max_{s}\left[e(s)-s(\mathcal{W}/t)\right]
\end{equation}
 maximising over the conjugate variable $s$. Consequently using the symmetry relation of equation (\ref{GCsym}) we may write
\begin{align}
\hat{e}(\mathcal{W}/t) & =\max_{s}\left[e(s)-(1-s)(\mathcal{W}/t)\right]\nonumber \\
 & =\max_{s}\left[e(s)+s(\mathcal{W}/t)\right]-(\mathcal{W}/t)\nonumber \\
 & =\hat{e}(-\mathcal{W}/t)-(\mathcal{W}/t).
\end{align}
 Since we expect large deviation behaviour described by equation (\ref{largedev}) this implies
\begin{equation}
{P}(\mathcal{W}/t)\simeq{ P}(-\mathcal{W}/t){\rm e}^{\mathcal{W}}
\end{equation}
or equivalently
\begin{equation}
{ P}(\mathcal{W})\simeq{ P}(-\mathcal{W}){\rm e}^{\mathcal{W}}
\end{equation}
 which is clearly analogous to the fluctuation theorems we have seen
previously. Taking a closer look at the action functional $\mathcal{W}$
we see that it is, for the systems we have been considering, a representation
of the entropy produced in the medium or a measure of the heat dissipated,
up to a constant $k_{B}$. Unlike the fluctuation theorems we considered
earlier, this is not guaranteed for all systems. To get a basic
understanding of this subtlety, we write the asymptotic fluctuation
theorem for the medium entropy production for the continuous systems
we have been considering in the form:
\begin{equation}
\frac{{ P}(\Delta S_{{\rm med}})}{{ P}(-\Delta S_{{\rm med}})}\simeq{\rm e}^{\Delta S_{{\rm med}}/k_{B}}.
\end{equation}
 However, we know that when considering steady states the following
fluctuation theorem holds for all time
\begin{equation}
\frac{{ P}(\Delta S_{{\rm tot}})}{{ P}(-\Delta S_{{\rm tot}})}={\rm e}^{\Delta S_{{\rm tot}}/k_{B}}.
\end{equation}
 Since $\Delta S_{{\rm tot}}=\Delta S_{{\rm med}}+\Delta S_{{\rm sys}}$
we may understand that the asymptotic symmetry will exist
when the system entropy change is negligible compared to the medium
entropy change. In steady states we expect a continuous dissipation
of heat and thus increase of medium entropy along with a change in
system entropy that \emph{on average} is zero. One might naively suggest
that this guarantees the asymptotic symmetry since the medium entropy
is unbounded and can grow indefinitely. This however, is not a strong
enough condition since if the system configuration space is unbounded
one cannot in general rule out large fluctuations to regions with
\emph{arbitrarily} low probability densities and therefore large changes
in system entropy which in principle can persist on any timescale.
What one requires to guarantee such a relation is the ability to neglect,
in detail for all trajectories, the system entropy change compared
with medium entropy change on long timescales. One can do so in general
if we insist that the state space is bounded. This means that the
system entropy has a well defined maximum and minimum value which
can be assumed to be unimportant on long timescales and so the asymptotic
symmetry necessarily follows. We note finally that systems with unbounded
state space are ubiquitous and include simple harmonic oscillators \cite{shoextension} and so investigations of fluctuation
theorems for such systems can and have lead to a wealth of non-trivial generalisations
and extensions.

\subsection{Generalisations and consideration of alternative dynamics}

What we hope the reader might appreciate following reading this chapter
is the malleability of quantities which satisfy fluctuation relations.
It is not difficult to produce quantities which obey relations similar
to the fluctuation theorems (although it may be hard to
show that they have any particular physical relevance) since the procedure simply
relies on a transformation of a path integral utilising the definition
of the entropy production itself. To clarify this point we consider
some generalisations of the relations we have seen. Let us consider
a new quantity which has the same form as the total entropy production
\begin{equation}
G[X]=k_{B}\ln{\left[\frac{P^{F}[X]}{P[Y]}\right]}
\end{equation}
 Here $P^{F}[X]$ is the same as before, yet we deliberately say very
little about $P[Y]$ other than it is a probability density of observing some
path $Y$ related to $X$ defined on the same space as $X$. Let us
compute the average of the negative exponential of this quantity
\begin{align}
\langle{\rm e}^{-G/k_{B}}\rangle & =\int dX\; P_{{\rm start}}(X(0))P^{F}[X(\tau)|X(0)]\frac{P(Y(0))P[Y(\tau)|Y(0)]}{P_{{\rm start}}(X(0))P^{F}[X(\tau)|X(0)]}\nonumber \\
 & =\int dY\; P(Y(0))P[Y(\tau)|Y(0)]\nonumber \\
 & =1\label{generalised}
\end{align}
 As long as $dX=dY$ and the unspecified initial distribution $P(Y(0))$
and transition probability density $P[Y(\tau)|Y(0)]$ are normalised then
\emph{any} such quantity $G[X]$ will obey an integral fluctuation
theorem averaged over the forward process. Clearly there are as many
relations as there are ways to define $P[Y(\tau)|Y(0)]$  and $P(Y(0))$ and most
will mean very little at all \cite{shargel}. However there are several such relations
in the literature obtained by an appropriate choice of $P(Y(0))$
and $P[Y(\tau)|Y(0)]$ that say something meaningful including,
for example, the Seifert end-point relations  \cite{seifertendpoint}. We will very briefly
allude to just two ways that this can be achieved by first noting
that one may consider an alternative dynamics known as `adjoint' dynamics,
leading to conditional path probabilities written $P_{\rm ad}[Y(\tau)|Y(0)]$, defined
such that they generate the same stationary distribution as the normal
dynamics, but with a current of the opposite sign \cite{Jarpathintegral}. For the
overdamped motion that we have been considering, where $P^{F}[X]=P_{{\rm start}}(x(0))P^{F}[x(\tau)|x(0)]$,
we may consider
\begin{itemize}
\item \emph{Hatano-Sasa relation}:\\
 \\
 By choosing
\begin{equation}
P(Y(0))=P_{{\rm start}}^{R}(\bar{x}(0))
\end{equation}
 and
\begin{equation}
P[Y(\tau)|Y(0)]=P_{{\rm ad}}^{R}[\bar{x}(\tau)|\bar{x}(0)]
\end{equation}
 we obtain the Hatano-Sasa relation \cite{hatanosasa} or integral fluctiation theorem for the `non-adiabatic entropy
production'  \cite{adiabaticnonadiabatic0,adiabaticnonadiabatic1,adiabaticnonadiabatic2} which concerns the so-called `excess heat' transferred to the environment\cite{oono}
such that
\begin{equation}
\langle\exp{[-\Delta Q_{{\rm ex}}/k_{B}T-\Delta S_{{\rm sys}}/k_{B}]}\rangle=1.
\label{hatanosasa:eq}
\end{equation}
 The (negative) exponent here is best described as the entropy production
associated with movement to stationarity, which, phenomenologically
includes transitions between non-equilibrium stationary states for which it was first
derived. This use of the $P_{{\rm ad}}^{R}$ adjoint dynamics is frequently
described as a reversal of both the protocol and dynamics \cite{Jarpathintegral} to be contrasted
with reversal of just the protocol for the integral fluctuation theorem.\\

\item \emph{Relation for the housekeeping heat}:\\
 \\
 By choosing
\begin{equation}
P(Y(0))=P_{{\rm start}}({x}(0))
\end{equation}
 and
\begin{equation}
P[Y(\tau)|Y(0)]=P_{{\rm ad}}^{F}[{x}(\tau)|{x}(0)]
\end{equation}
 we obtain the Speck-Seifert integral relation  \cite{IFThousekeeping} or integral fluctuation theorem for the `adiabatic
entropy production'  \cite{adiabaticnonadiabatic0,adiabaticnonadiabatic1,adiabaticnonadiabatic2}, which concerns the so-called `housekeeping
heat' absorbed by the environment \cite{oono} such that
\begin{equation}
\langle\exp{[-\Delta Q_{{\rm hk}}/k_{B}T]}\rangle=1
\label{speckseifert:eq}
\end{equation}
 where $\Delta Q_{{\rm hk}}=\Delta Q_{{\rm med}}-\Delta Q_{{\rm ex}}$ \cite{oono}, and where the negative exponent
is best described as the entropy production or heat dissipation associated
with the non-equilibrium steady state. Such a consideration might be called a reversal
of the dynamics, but not the protocol.
\end{itemize}

\noindent Both of these relations are relevant to the study of systems where detailed balance does not hold and amount to a division in the total entropy production, or irreversibility, into the two types we considered in section \ref{detailedbalance:sec}, namely movement towards stationarity brought about by driving and relaxation, and the breaking of time reversal symmetry that arises specifically when detailed balance is absent. Consequently if detailed balance does hold then the exponent in equation (\ref{speckseifert:eq}) is zero and equation (\ref{hatanosasa:eq}) reduces to the integral fluctuation theorem.

\section{Fluctuation relations for reversible deterministic systems}
\label{determ}

So far we have chosen to focus on systems that obey stochastic dynamics,
whereby the interaction with the environment, and the explicit breakage
of time reversal symmetry, is implemented through the presence of random forces
in the equations of motion. However, there exists a framework for
deriving fluctuation relations that is based on deterministic dynamical
equations, whereby the environmental interaction is represented through
specific non-linear terms \cite{Evans02} which supplement the usual terms in Newton's equations. These have the effect of constraining
some aspect of the system, such as its kinetic energy, either to a
chosen constant, or to a particular distribution as time progresses.
Most importantly, they can be reversible, such that a trajectory driven
by a specified protocol, and its time-reversed counterpart driven
by a time-reversed protocol, are both solutions to the dynamics.
In practice these so-called thermostatting terms which provide the non-linearity are
taken to act solely on the boundaries (which can be made arbitrarily remote) in order for the system to be unaffected by the precise details of
the input and removal of heat. This
provides a parallel framework within which the dynamics of an open
system, and hence fluctuation relations, can be explored. Indeed it
was through the consideration of deterministic, reversible dynamical
systems that many of the seminal insights into fluctuation relations
were first obtained \cite{Evans93}.

Given that there was a choice over the framework to employ, we opted
to use stochastic dynamics to develop this pedagogical overview. This
has some benefit in that the concept of entropy change can readily
be attached to the idea of the growth of uncertainty in system evolution,
identifying it explicitly with the intrinsic irreversibility of the
stochastic dynamics. Nevertheless, it is important to review the deterministic
approach as well, and explore some of the additional insight that
it provides.

The main outcome of seminal and ongoing studies by Evans, Searles and coworkers \cite{Evans93,Evans94,Evans2011} is the identification of a system property that displays a tendency
to grow with time under specified non-Hamiltonian reversible dynamics.
The development of the $H$-theorem by Boltzmann was a similar attempt
to identify such a quantity. However, we shall have to confront the fact that
by their very nature, deterministic equations do not generate additional
uncertainty as time progresses. The configuration of a system at a
time $t$ is precisely determined given the configuration at $t_{0}$.
Even if the latter were specified only through a probability distribution,
all future and past configurations associated with each starting configuration
are fixed, and uncertainty is therefore not changed by the passage
of time. Something other than the increase in uncertainty will have
to emerge in a deterministic framework if it is to represent entropy
change.

Within such a framework, a system is described in terms of a probability density for its dynamical
variables $x,v,\cdots$, collectively denoted $\Gamma$. An initial
probability density $P(\Gamma,0)$ evolves under the dynamics into
a density $P(\Gamma,$t). Furthermore, the starting `point' of a trajectory
$\Gamma_{0}$ (namely ($x(0),v(0)\cdots$) is linked uniquely to a
terminating point $\Gamma_{t}$ (namely $x(t),v(t),\cdots$), passing
through points $\Gamma_{t^{\prime}}$ in between. It may then be shown
\cite{Evans93,Evans94} that
\begin{equation}
P(\Gamma_{t},t)=P(\Gamma_{0},0)\exp\left(-\int_{0}^{t}\Lambda(\Gamma_{t^{\prime}})dt^{\prime}\right),\label{eq:201}
\end{equation}
where $\Lambda(\Gamma_{t^{\prime}})$ is known as the phase space
contraction rate associated with configuration $\Gamma_{t^{\prime}}$,
which may be related specifically to the terms in the equations of
motion that impose the thermal constraint. For a system without constraint,
and hence thermally isolated, the phase space contraction rate is
therefore zero everywhere, and the resulting $P(\Gamma_{t},t)=P(\Gamma_{0},0)$
is an expression of Liouville's theorem: the conservation of probability
density along any trajectory followed by the system.

For typically employed thermal constraints (denoted thermostats/ergostats,
depending on their nature) it may be shown that the phase space contraction
rate is related to the rate of heat transfer to the system from the
implied environment. For the so-called Nose-Hoover thermostat
at fixed target temperature $T$ we are able to write $\int_{0}^{t}\Lambda(\Gamma_{t^{\prime}})dt^{\prime}=\Delta Q(\Gamma_{0})/kT$,
where $\Delta Q(\Gamma_{0})$ is the heat transferred to the system
over the course of a trajectory of duration $t$ starting from $\Gamma_{0}$.

We now consider the dissipation function $\Omega(\Gamma)$ defined
through
\begin{equation}
\int_{0}^{t}\Omega(\Gamma_{t^{\prime}})dt^{\prime}=\ln\left(\frac{P(\Gamma_{0},0)}{P(\Gamma_{t}^{*},0)}\exp\left(-\int_{0}^{t}\Lambda(\Gamma_{t^{\prime}})dt^{\prime}\right)\right),\label{eq:201a}
\end{equation}
where $\Gamma_{t}^{*}$ is related to $\Gamma_{t}$ by the reversal
of all velocity coordinates. Assuming that the probability density
at time zero is symmetric in velocities, such that $P(\Gamma_{0},0)=P(\Gamma_{0}^{*},0)$
(which ensures that the right hand side of equation (\ref{eq:201a})
vanishes when $t=0$), we can write
\begin{equation}
\int_{0}^{t}\Omega(\Gamma_{t^{\prime}})dt^{\prime}=\bar{\Omega}_{t}(\Gamma_{0})t=\ln\left(\frac{P(\Gamma_{t},t)}{P(\Gamma_{t},0)}\right),\label{eq:202}
\end{equation}
defining a mean dissipation function $\bar{\Omega}_{t}(\Gamma_{0})$
for the trajectory starting from $\Gamma_{0}$ and of duration $t$.
It is a quantity that will take a variety of values for a given protocol
(the specified dynamics over the period in question), depending on $\Gamma_0$, and its distribution
has particular properties, just as we found for the distributions
of functionals such as $\Delta S_{{\rm tot}}$ in equation (\ref{totalent}).
For example, we have
\begin{eqnarray}
\langle\exp\left(-\bar{\Omega}_{t}t\right)\rangle & = & \int d\Gamma_{0}P(\Gamma_{0},0)\exp\left(-\bar{\Omega}_{t}(\Gamma_{0})t\right)\nonumber \\
 & = & \int d\Gamma_{t}P(\Gamma_{t},t)\exp\left(-\bar{\Omega}_{t}t\right)\nonumber \\
 & = & \int d\Gamma_{t}P(\Gamma_{t},0)=1,\label{eq:203}
\end{eqnarray}
where the averaging is over the various possibilities for $\Gamma_{0}$, or equivalently $\Gamma_t$,
and where we have imposed a probability conservation condition $d\Gamma_{0}P(\Gamma_{0},0)=d\Gamma_{t}P(\Gamma_{t},t)$,
implying that $d\Gamma_{t}$ is the region of phase space around $\Gamma_{t}$
that contains all the end-points of trajectories starting within the
region $d\Gamma_{0}$ around $\Gamma_{0}$. This result takes the
same form as the integral fluctuation theorem obtained using stochastic
dynamics, but now involves the mean dissipation function. In the deterministic
dynamics literature, the result equation (\ref{eq:203}) is known
as a non-equilibrium partition identity. As a consequence, we
can deduce that $\langle\bar{\Omega}_{t}\rangle\ge0$, again a result
that resembles several already encountered. 

Now let us consider a protocol that is time-symmetric about its midpoint,
and for simplicity, consists of time variation in the form of the
system Hamiltonian. The thermal constraint, as discussed above, is
imposed through reversible non-Hamiltonian terms in the dynamics (let us say the Nose-Hoover scheme) and is explicitly time-independent
and therefore isothermal. For such a case it is clear that a trajectory running
from $\Gamma_{0}$ to $\Gamma_{t}$ over the time period $0\to t$
can be generated in a velocity-reversed form, and in reverse sequence,
by evolving for the same period forwards in time under the same equations
of motion, but starting from the velocity-reversed configuration at
time $t$, namely $\Gamma_{t}^{*}$. This evolution is precisely that
which would be obtained by running a movie of the normal trajectory
backwards. The velocity-reversed or time reversed counterpart to each
phase space point $\Gamma_{t^{\prime}}$ is visited, but in the opposite
order, and the final configuration is $\Gamma_{0}^{*}$. The mean
dissipation function for such a trajectory would be given by
\begin{equation}
\bar{\Omega}_{t}(\Gamma_{t}^{*})t=\ln\left(\frac{P(\Gamma_{t}^{*},0)}{P(\Gamma_{0},0)}\exp\left(-\Delta Q(\Gamma_{t}^{*})/kT\right)\right),\label{eq:203a}
\end{equation}
where $\Delta Q(\Gamma_{t}^{*})$ is the heat transfer for this time-reversed
trajectory. The symmetry of the protocol, and the symmetry of the
Hamiltonian under velocity reversal, allows us to conclude that the
heat transfer associated with the trajectory starting from $\Gamma_{0}$
is equal and opposite to that associated with starting point $\Gamma_{t}^{*}$,
and hence that the mean dissipation functions for the two trajectories
must satisfy $\bar{\Omega}_{t}(\Gamma_{0})=-\bar{\Omega}_{t}(\Gamma_{t}^{*})$.
We can then proceed to derive a specific case of the\\\\
 \emph{Evans-Searles Fluctuation Theorem} (ESFT) \\
\\
associated with the distribution of values
$\bar{\Omega}_{t}$ taken by the mean dissipation function $\bar{\Omega}_{t}(\Gamma_{0})$:
\begin{eqnarray}
P(\bar{\Omega}_{t}) & = & \int d\Gamma_{0}P(\Gamma_{0},0)\delta\left(\bar{\Omega}_{t}(\Gamma_{0})-\bar{\Omega}_{t}\right)=\int d\Gamma_{t}P(\Gamma_{t},t)\delta\left(\bar{\Omega}_{t}(\Gamma_{0})-\bar{\Omega}_{t}\right)\nonumber \\
 & = & \int d\Gamma_{t}P(\Gamma_{t},t)\exp\left(\bar{\Omega}_{t}(\Gamma_{0})t\right)\frac{P(\Gamma_{t},0)}{P(\Gamma_{t},t)}\delta\left(\bar{\Omega}_{t}(\Gamma_{0})-\bar{\Omega}_{t}\right)\nonumber \\
 & = & \exp\left(\bar{\Omega}_{t}t\right)\int d\Gamma_{t}P(\Gamma_{t},0)\delta\left(\bar{\Omega}_{t}(\Gamma_{0})-\bar{\Omega}_{t}\right)\nonumber \\
 & = & \exp\left(\bar{\Omega}_{t}t\right)\int d\Gamma_{t}P(\Gamma_{t},0)\delta\left(-\bar{\Omega}_{t}(\Gamma_{t}^{*})-\bar{\Omega}_{t}\right)\nonumber \\
 & = & \exp\left(\bar{\Omega}_{t}t\right)\int d\Gamma_{t}^{*}P(\Gamma_{t}^{*},0)\delta\left(-\bar{\Omega}_{t}(\Gamma_{t}^{*})-\bar{\Omega}_{t}\right)\nonumber \\
 & = & \exp\left(\bar{\Omega}_{t}t\right)P(-\bar{\Omega}_{t}),\label{eq:204}
\end{eqnarray}
noting that the Jacobian for the transformation of the integration
variables from $\Gamma_{t}$ to $\Gamma_{t}^{*}$ is unity. Under
the assumed conditions, therefore, we have obtained a relation that
resembles (but historically preceded) the transient fluctuation theorem equation
(\ref{DFTSeif}) or detailed fluctuation theorem equation (\ref{DFTgen}) derived within
the framework of stochastic dynamics. It only remains to make connections
between the mean dissipation function and thermodynamic quantities
to complete the parallel development, though it has been argued that the mean dissipation function itself is the
more general measure of non-equilibrium behaviour \cite{Evans2011}.

If we assume that the initial distribution is one of canonical equilibrium,
such that $P(\Gamma_{0},0)\propto\exp\left(-H(\Gamma_{0},0)/kT\right)$,
where $H(\Gamma_{0},0)$ is the system Hamiltonian at $t=0$, then
we find from equation (\ref{eq:201a}) that
\begin{equation}
\bar{\Omega}_{t}(\Gamma_{0})t=\frac{1}{kT}\left(H(\Gamma_{t},0)-H(\Gamma_{0},0)\right)-\frac{1}{kT}\Delta Q(\Gamma_{0}),\label{eq:205}
\end{equation}
and if the Hamiltonian at time $t$ takes the same functional form
as it does at $t=0$, then $H(\Gamma_{t},0)=H(\Gamma_{t},t)$ and
we get
\begin{eqnarray}
\bar{\Omega}_{t}(\Gamma_{0})t & = & \frac{1}{kT}\left(H(\Gamma_{t},t)-H(\Gamma_{0},0)\right)-\frac{1}{kT}\Delta Q(\Gamma_{0})\label{eq:206}\\
 & = & \frac{1}{kT}\left(\Delta E(\Gamma_{0})-\Delta Q(\Gamma_{0})\right)=\frac{1}{kT}\Delta W(\Gamma_{0}),
\end{eqnarray}
where $\Delta E(\Gamma_{0})$ is the change in system energy along
a trajectory starting from $\Gamma_{0}$. Hence the mean dissipation
function is proportional to the (here solely thermodynamic) work performed
on the system as it follows the trajectory starting from $\Gamma_{0}$. We deduce that the expectation value
of this work is positive, and that the probability distribution of
work for a time-symmetric protocol, and starting from canonical equilibrium,
satisfies the ESFT.

Deterministic methods may be used to derive a variety of statistical
results involving the work performed on a system, including the Jarzynski
equation and the Crooks relation. Non-conservative work may be included
such that relations analogous to equation (\ref{FRmech}) may be obtained.
However, it seems that a parallel development of the statistics of
$\Delta S_{{\rm tot}}$ is not possible. The fundamental problem is
revealed if we try to construct the deterministic counterpart to the
total entropy production defined in equations (\ref{totalent}) and (\ref{ratio1}):
\begin{equation}
\Delta S_{{\rm tot}}^{{\rm det}}=\Delta S_{{\rm sys}}+\Delta S_{{\rm med}}=-k\ln\left(\frac{P(\Gamma_{t},t)}{P(\Gamma_{0},0)}\right)-\frac{\Delta Q(\Gamma_{0})}{T}.\label{eq:207}
\end{equation}
According to equation (\ref{eq:201}) this is identically zero.
As might have been expected, uncertainty does not change under deterministic
dynamics, and total entropy, in the form that we have chosen to define
it, is constant. Nevertheless, the derivation of relationships involving
the statistics of work performed and heat transferred, just alluded to, corresponding
to similar expressions obtained using the stochastic dynamics framework,
indicate that the use of deterministic reversible dynamics is an equivalent
procedure for describing the behaviour.
Pedagogically it is perhaps best to focus on just one approach, but
a wider appreciation of the field requires an awareness of both.

\section{Examples of the fluctuation relations in action}

The development of theoretical results of the kind we have seen so
far is all very well, but their meaning is best appreciated by considering
examples, which we do in this section. We shall consider overdamped
stochastic dynamics, such that the velocities are always in an equilibrium
Maxwell-Boltzmann distribution and never enter into consideration
for entropy production. And in the first two cases we shall focus
on the harmonic oscillator, since we understand its properties well.
The only drawback of the harmonic oscillator is that it is a rather
special case and some of its properties are not general \cite{shargel,Saha09} though we deliberately avoid situations where the distributions produced are gaussian in these examples. In the third
case we describe the simplest of non-equilibrium steady states and
illustrate a detailed fluctuation theorem for the entropy, and its origin in
the breaking of detailed balance.

\subsection{Harmonic oscillator subject to a step change in spring constant}

Let us form the most simple model of the compression-expansion type
processes that are ubiquitous within thermodynamics. We start by considering
a 1-d classical harmonic oscillator subject to a Langevin heat bath.
Such a system is governed by the overdamped equation of motion
\begin{equation}
\dot{x}=-\frac{\kappa x}{m\gamma}+\left(\frac{2k_{B}T}{m\gamma}\right)^{1/2}\xi(t),
\end{equation}
 where $\kappa$ is the spring constant. The corresponding Fokker-Planck
equation is
\begin{equation}
\frac{\partial P(x,t)}{\partial t}=\frac{1}{m\gamma}\frac{\partial\left(\kappa xP(x,t)\right)}{\partial x}+\frac{k_{B}T}{m\gamma}\frac{\partial^{2}P(x,t)}{\partial x^{2}}.\label{eq:126}
\end{equation}
 We consider a simple work process whereby, starting from equilibrium
at temperature $T$, we instigate an instantaneous step change in
spring constant from $\kappa_{0}$ to $\kappa_{1}$ at $t=0$ with
the system in contact with the thermal bath at all times. This has
the effect of compressing or expanding the distribution. We are then
interested in the statistics of the entropy change associated with
the process. Starting from equations (\ref{totalent}) and (\ref{ratio1}) for our definition of
the entropy production we may write
\begin{equation}
\Delta S_{{\rm tot}}=\frac{\Delta W-\Delta\phi}{T}+k_{B}\ln\left(\frac{P_{{\rm start}}(x_{0})}{P_{{\rm end}}(x_{1})}\right),\label{eq:89}
\end{equation}
 utilising notation $x_{1}=x(t)$ and $x_{0}=x(0)$ and $\phi(x)=\kappa x^2/2$. We also have
\begin{equation}
\Delta W=\frac{1}{2}\left(\kappa_{1}-\kappa_{0}\right)x_{0}^{2},
\end{equation}
 and
\begin{equation}
\Delta\phi=\frac{1}{2}\kappa_{1}x_{1}^{2}-\frac{1}{2}\kappa_{0}x_{0}^{2},
\end{equation}
 and so can write
\begin{equation}
\Delta S_{{\rm tot}}=-\frac{\kappa_{1}}{2T}\left(x_{1}^{2}-x_{0}^{2}\right)+k_{B}\ln\left(\frac{P_{{\rm start}}(x_{0})}{P_{{\rm end}}(x_{1})}\right)
\label{stotosc}
\end{equation}
 Employing an initial canonical distribution
\begin{equation}
P_{{\rm start}}(x_{0})=\left(\frac{\kappa_{0}}{2\pi k_{B}T}\right)^{1/2}\exp\left(-\frac{\kappa_{0}x_{0}^{2}}{2k_{B}T}\right),\label{eq:90}
\end{equation}
 the distribution at the end of the process will be given by
\begin{equation}
P_{{\rm end}}(x_{1})=\int_{-\infty}^{\infty}dx_{0}P_{{\rm OU}}^{\kappa_{1}}[x_{1}|x_{0}]P_{{\rm start}}(x_{0}),\label{eq:92}
\end{equation}
 This is the Ornstein-Uhlenbeck process and so has transition probability
density $P_{\mathrm{OU}}^{\kappa}$ given by analogy to equation (\ref{eq:41}).
Hence we may write
\begin{align}
P_{{\rm end}}(x_{1}) & =\int_{-\infty}^{\infty}dx_{0}\left(\frac{\kappa_{1}}{2\pi k_{B}T\left(1\!-\!{\mathrm{e}}^{-\frac{2\kappa_{1}t}{m\gamma}}\right)}\right)^{1/2}\exp\left(-\frac{\kappa_{1}\big(x_{1}-x_{0}{\mathrm{e}}^{-\frac{\kappa_{1}t}{m\gamma}}\big)^{2}}{2k_{B}T\left(1\!-\!{\mathrm{e}}^{-\frac{2\kappa_{1}t}{m\gamma}}\right)}\right)\nonumber \\
 & \qquad\times\left(\frac{\kappa_{0}}{2\pi k_{B}T}\right)^{1/2}\exp\left(-\frac{\kappa_{0}x_{0}^{2}}{2k_{B}T}\right)\nonumber \\
 & =\left(\frac{\tilde{\kappa}(t)}{2\pi k_{B}T}\right)^{1/2}\exp\left(-\frac{\tilde{\kappa}(t)x_{1}^{2}}{2k_{B}T}\right),\label{eq:93}
\end{align}
 where
\begin{equation}
\tilde{\kappa}(t)=\frac{\kappa_{0}\kappa_{1}}{\kappa_{0}+{\rm e}^{-2\kappa_{1}t/(m\gamma)}\left(\kappa_{1}-\kappa_{0}\right)},\label{eq:94}
\end{equation}
 such that $P_{{\rm end}}$ is always gaussian. The coefficient $\tilde{\kappa}(t)$
evolves monotonically from $\kappa_{0}$ at $t=0$ to $\kappa_{1}$
as $t\rightarrow\infty$. Substituting this into equation (\ref{stotosc}) allows us to
write
\begin{equation}
\Delta S_{{\rm tot}}(x_{1},x_{0},t)=-\frac{\kappa_{1}}{2T}\left(x_{1}^{2}-x_{0}^{2}\right)+\frac{k_{B}}{2}\ln\left(\frac{\kappa_{0}}{\tilde{\kappa}(t)}\right)-\frac{\kappa_{0}x_{0}^{2}}{2T}+\frac{\tilde{\kappa}(t)x_{1}^{2}}{2T},\label{eq:95}
\end{equation}
 for the entropy production associated with a trajectory that begins
at $x_{0}$ and ends at $x_{1}$ at time $t$, and is not specified
in between. We can average this over the probability distribution
for such a trajectory to get
\begin{align}
\langle\Delta S_{{\rm tot}}\rangle & =\int dx_{0}dx_{1}P_{{\rm OU}}^{\kappa_{1}}[x_{1}|x_{0}]P_{{\rm start}}(x_{0})\Delta S_{{\rm tot}}(x_{1},x_{0},t)\nonumber \\
 & =k_{B}\left(-\frac{1}{2}\frac{\kappa_{1}}{\tilde{\kappa}(t)}+\frac{1}{2}\frac{\kappa_{1}}{\kappa_{0}}+\frac{1}{2}\ln\left(\frac{\kappa_{0}}{\tilde{\kappa}(t)}\right)-\frac{1}{2}+\frac{1}{2}\right)\nonumber \\
 & =\frac{k_{B}}{2}\left(\frac{\kappa_{1}}{\kappa_{0}}-\frac{\kappa_{1}}{\tilde{\kappa}(t)}+\ln\left(\frac{\kappa_{0}}{\tilde{\kappa}(t)}\right)\right),\label{eq:96}
\end{align}
 making full use of the separation of $\Delta S_{{\rm tot}}$ into
quadratic terms, and the gaussian character of the distributions.
At $t=0$ $\langle\Delta S_{\rm tot}\rangle$ is zero, and as $t\rightarrow\infty$ we find
\begin{equation}
\lim_{t\to\infty}\langle\Delta S_{{\rm tot}}\rangle=\frac{k_{B}}{2}\left(\frac{\kappa_{1}}{\kappa_{0}}-1-\ln\left(\frac{\kappa_{1}}{\kappa_{0}}\right)\right),\label{eq:97}
\end{equation}
 which is positive since $\ln z\le z-1$ for all $z$. Furthermore,
\begin{equation}
\frac{d\langle\Delta S_{{\rm tot}}\rangle}{dt}=\frac{k_{B}}{2\tilde{\kappa}^{2}}\frac{d\tilde{\kappa}}{dt}\left(\kappa_{1}-\tilde{\kappa}\right),\label{eq:98}
\end{equation}
 and it is clear that this is positive at all times during the evolution,
irrespective of the values of $\kappa_{1}$ and $\kappa_{0}$. If
$\kappa_{1}>\kappa_{0}$ then $\tilde{\kappa}$ increases with time, whilst remaining less than $\kappa_1$, and all
factors on the right hand side of equation (\ref{eq:98}) are positive.
If $\kappa_{1}<\kappa_{0}$ then $\tilde{\kappa}$ decreases but always
remains greater than $\kappa_{1}$ and the mean total entropy production
is still positive as the relaxation process proceeds.

The work done on the system is simply the input of potential energy
associated with the shift in spring constant:
\begin{equation}
\Delta W(x_{1},x_{0},t)=\frac{1}{2}\left(\kappa_{1}-\kappa_{0}\right)x_{0}^{2},\label{eq:99}
\end{equation}
 and so the mean work performed up to any time $t>0$ is
\begin{equation}
\langle\Delta W\rangle=\frac{k_{B}T}{2}\left(\frac{\kappa_{1}}{\kappa_{0}}-1\right),\label{eq:100}
\end{equation}
 which is greater than $\Delta F=(k_{B}T/2)\ln\left(\kappa_{1}/\kappa_{0}\right)$.
The mean dissipative work is
\begin{equation}
\langle\Delta W_{d}\rangle=\langle\Delta W\rangle-\Delta F=\frac{k_{B}T}{2}\left(\frac{\kappa_{1}}{\kappa_{0}}-1-\ln\left(\frac{\kappa_{1}}{\kappa_{0}}\right)\right),\label{eq:101}
\end{equation}
 and this equals the mean entropy generated as $t\rightarrow\infty$
derived in equation (\ref{eq:97}), which is to be expected since the system
started in equilibrium. More specifically, let us verify the Jarzynski
equality:

\begin{align}
\langle\exp\left(-\Delta W/k_{B}T\right)\rangle & =\int dx_{0}P_{{\rm start}}(x_{0})\exp\left(-\left(\kappa_{1}-\kappa_{0}\right)x_{0}^{2}/2k_{B}T\right)\nonumber \\
 & =\left(\kappa_{0}/\kappa_{1}\right)^{1/2}=\exp\left(-\Delta F/k_{B}T\right),\label{eq:102}
\end{align}
 as required.

Now we demonstrate that the integral fluctuation relation is satisfied.
We consider
\begin{align}
\langle & \exp\left(-\Delta S_{{\rm tot}}/k_{B}\right)\rangle=\left\langle \exp\left(\frac{\kappa_{1}}{2k_{B}T}\left(x_{1}^{2}-x_{0}^{2}\right)-\frac{1}{2}\ln\left(\frac{\kappa_{0}}{\tilde{\kappa}}\right)+\frac{\kappa_{0}x_{0}^{2}}{2k_{B}T}-\frac{\tilde{\kappa}x_{1}^{2}}{2k_{B}T}\right)\right\rangle \nonumber \\
 & =\int dx_{1}dx_{0}P_{{\rm OU}}^{\kappa_{1}}[x_{1}|x_{0}]P_{{\rm start}}(x_{0})\nonumber \\
 & \qquad\times\exp\left(\frac{\kappa_{1}}{2k_{B}T}\left(x_{1}^{2}-x_{0}^{2}\right)-\frac{1}{2}\ln\left(\frac{\kappa_{0}}{\tilde{\kappa}}\right)+\frac{\kappa_{0}x_{0}^{2}}{2k_{B}T}-\frac{\tilde{\kappa}x_{1}^{2}}{2k_{B}T}\right)\nonumber \\
 & =\left(\frac{\tilde{\kappa}}{\kappa_{0}}\right)^{1/2}\int dx_{1}dx_{0}\bigg(\frac{\kappa_{1}}{2\pi k_{B}T\Big(1-{\mathrm{e}}^{-\frac{2\kappa_{1}t}{m\gamma}}\Big)}\bigg)^{1/2}\exp{\left(-\frac{\kappa_{1}\big(x_{1}-x_{0}{\rm e}^{-\frac{\kappa_{1}t}{m\gamma}}\big)^{2}}{2k_{B}T\left(1-{\mathrm{e}}^{-\frac{2\kappa_{1}t}{m\gamma}}\right)}\right)}\nonumber \\
 & \qquad\times\left(\frac{\kappa_{0}}{2\pi k_{B}T}\right)^{1/2}\exp\left(-\frac{\kappa_{0}x_{0}^{2}}{2k_{B}T}\right)\exp\left(\frac{\kappa_{1}}{2k_{B}T}\left(x_{1}^{2}-x_{0}^{2}\right)+\frac{\kappa_{0}x_{0}^{2}}{2k_{B}T}-\frac{\tilde{\kappa}x_{1}^{2}}{2k_{B}T}\right)\nonumber \\
 & =1,\label{eq:102a}
\end{align}
 which is a tedious integration, but the result is inevitable.

Further we can directly affirm the Crooks relation for this process.
The work over the forward process is given by equation (\ref{eq:99})
and so, choosing $\kappa_1> \kappa_0$, we can relate its distribution according to
\begin{align}
{ P}^{F}(\Delta W) & =P_{\rm eq}(x_{0})\frac{dx_{0}}{d\Delta W}\nonumber \\
 & =\left(\frac{\kappa_{0}}{2\pi k_{B}T}\right)^{1/2}\exp\left(-\frac{\kappa_{0}x_{0}^{2}}{2k_{B}T}\right)\frac{1}{\kappa_{1}-\kappa_{0}}\left(\frac{2\Delta W}{\kappa_{1}-\kappa_{0}}\right)^{-1/2}H(\Delta W)\nonumber \\
 & =\frac{1}{\sqrt{4\pi k_{B}T}}\left(\frac{\kappa_{0}}{\kappa_{1}-\kappa_{0}}\right)^{1/2}\Delta W^{-1/2}\exp{\left(-\frac{\kappa_{0}}{\kappa_{1}-\kappa_{0}}\frac{\Delta W}{k_{B}T}\right)}H(\Delta W)
\label{forwork}
\end{align}
 where $H(\Delta W)$ is the Heaviside step function. Let us consider the appropriate reverse process. Starting in equilibrium
defined by $\kappa_{1}$, where again to form the reversed protocol we must have $\kappa_1>\kappa_0$, the work is
\begin{equation}
\Delta W=\frac{1}{2}\left(\kappa_{0}-\kappa_{1}\right)x_{1}^{2}
\end{equation}
 and so we can relate its distribution according to
\begin{align}
{ P}^{R}(\Delta W) & =P_{\rm eq}(x_{1})\frac{dx_{1}}{d\Delta W}\nonumber \\
 & =\left(\frac{\kappa_{1}}{2\pi k_{B}T}\right)^{1/2}\exp\left(-\frac{\kappa_{1}x_{1}^{2}}{2k_{B}T}\right)\frac{1}{\kappa_{0}-\kappa_{1}}\left(\frac{2\Delta W}{\kappa_{0}-\kappa_{1}}\right)^{-1/2}H(-\Delta W)\nonumber \\
 & =\frac{1}{\sqrt{4\pi k_{B}T}}\left(\frac{\kappa_{1}}{\kappa_{0}-\kappa_{1}}\right)^{1/2}\Delta W^{-1/2}\exp{\left(-\frac{\kappa_{1}}{\kappa_{0}-\kappa_{1}}\frac{\Delta W}{k_{B}T}\right)}H(-\Delta W)
\end{align}
 so that
\begin{align}
{ P}^{R}(-\Delta W) & =\frac{1}{\sqrt{4\pi k_{B}T}}\left(\frac{\kappa_{1}}{\kappa_{0}-\kappa_{1}}\right)^{1/2}(-\Delta W)^{-1/2}\exp{\left(\frac{\kappa_{1}}{\kappa_{0}-\kappa_{1}}\frac{\Delta W}{k_{B}T}\right)}H(\Delta W)\nonumber \\
 & =\frac{1}{\sqrt{4\pi k_{B}T}}\left(\frac{\kappa_{1}}{\kappa_{1}-\kappa_{0}}\right)^{1/2}(\Delta W)^{-1/2}\exp{\left(-\frac{\kappa_{1}}{\kappa_{1}-\kappa_{0}}\frac{\Delta W}{k_{B}T}\right)}H(\Delta W)
 \label{backwork}
\end{align}
 Taking the ratio of these two distributions (\ref{forwork}) and (\ref{backwork}) gives
\begin{align}
\frac{{ P}^{F}(\Delta W)}{{ P}^{R}(-\Delta W)} & =\sqrt{\frac{\kappa_{0}}{\kappa_{1}}}\exp{\left(-\frac{\kappa_{0}-\kappa_{1}}{\kappa_{1}-\kappa_{0}}\frac{\Delta W}{k_{B}T}\right)}\nonumber \\
 & =\exp{\left(\frac{\Delta W-\frac{k_{B}T}{2}\ln{\frac{\kappa_{1}}{\kappa_{0}}}}{k_{B}T}\right)}\nonumber \\
 & =\exp{\left(\frac{\Delta W-\Delta F}{k_{B}T}\right)}
\end{align}
 which is the Crooks work relation as required.

\subsection{Smoothly squeezed harmonic oscillator}

Now let us consider a process where work is performed isothermally
on a particle, but this time by a continuous variation of the spring
constant. We have
\begin{equation}
\Delta W=\int_{0}^{\tau}\frac{\partial\phi(x(t),\lambda(t))}{\partial\lambda}\frac{d\lambda}{dt}dt,
\end{equation}
 where $\lambda(t)=\kappa(t)$ and $\phi(x(t),\kappa(t))=\frac{1}{2}\kappa(t)x_{t}^{2}$,
where $x_{t}=x(t)$, such that
\begin{equation}
\Delta W=\int_{0}^{\tau}\frac{1}{2}\dot{\kappa}(t)x_{t}^{2}dt.
\end{equation}
 Similarly the change in system energy will be given simply
by
\begin{equation}
\Delta\phi=\int_{0}^{\tau}\frac{d\phi(x(t),\lambda(t))}{dt}dt=\frac{1}{2}\kappa(\tau)x_{\tau}^{2}-\frac{1}{2}\kappa_{0}x_{0}^{2}.
\end{equation}
 Accordingly we can once again describe the entropy production as
\begin{equation}
\Delta S_{{\rm tot}}=\frac{1}{2T}\int_{0}^{\tau}\dot{\kappa}x_{t}^{2}dt-\frac{1}{2T}\kappa(\tau)x_{\tau}^{2}+\frac{1}{2T}\kappa_{0}x_{0}^{2}+k_{B}\ln\left(\frac{P_{{\rm start}}(x_{0})}{P_{{\rm end}}(x_{\tau})}\right).\label{eq:103}
\end{equation}
 For convenience, we assume the initial state to be in
canonical equilibrium. The evolving distribution $P$ satisfies the
appropriate Fokker-Planck equation:
\begin{equation}
\frac{\partial P}{\partial t}=\frac{\kappa(t)}{m\gamma}\frac{\partial\left(xP\right)}{\partial x}+\frac{k_{B}T}{m\gamma}\frac{\partial^{2}P}{\partial x^{2}}.\label{eq:104}
\end{equation}
Since $P$ is initially canonical, it retains its gaussian form and
can be written
\begin{equation}
P_{{\rm end}}(x_{\tau})=P(x_{\tau},\tau)=\left(\frac{\tilde{\kappa}(\tau)}{2\pi k_{B}T}\right)^{1/2}\exp\left(-\frac{\tilde{\kappa}(\tau)x_{\tau}^{2}}{2k_{B}T}\right),\label{eq:105}
\end{equation}
 where $\tilde{\kappa}(t)$ has its own equation of motion according
to
\begin{equation}
\frac{d\tilde{\kappa}}{dt}=-\frac{2}{m\gamma}\tilde{\kappa}\left(\tilde{\kappa}-\kappa\right),\label{eq:106}
\end{equation}
 with initial condition $\tilde{\kappa}(0)=\kappa_{0}$. We can solve
for $\tilde{\kappa}$: write $z=\tilde{\kappa}^{-1}$ such that
\begin{equation}
\frac{dz}{dt}=\frac{2}{m\gamma}\left(1-\kappa z\right).\label{eq:107}
\end{equation}
 This has integrating factor solution
\begin{equation}
z(\tau)\exp\left(\frac{2}{m\gamma}\int_{0}^{\tau}\kappa(t)dt\right)=z(0)+\int_{0}^{\tau}\exp\left(\frac{2}{m\gamma}\int_{0}^{t}\kappa(t^{\prime})dt^{\prime}\right)\frac{2}{m\gamma}dt,\label{eq:108}
\end{equation}
 or equivalently
\begin{equation}
\frac{1}{\tilde{\kappa}(\tau)}=\frac{1}{\kappa(0)}\exp\left(-\frac{2}{m\gamma}\int_{0}^{\tau}\kappa(t)dt\right)+\frac{2}{m\gamma}\int_{0}^{\tau}\exp\left(-\frac{2}{m\gamma}\int_{t}^{\tau}\kappa(t^{\prime})dt^{\prime}\right)dt.\label{eq:109}
\end{equation}
 Returning to the entropy production, we now write
\begin{equation}
\Delta S_{{\rm tot}}=\frac{1}{T}\int_{0}^{\tau}\frac{1}{2}\dot{\kappa}x_{t}^{2}dt-\frac{1}{2T}\kappa(\tau)x_{\tau}^{2}+\frac{1}{2T}\kappa_{0}x_{0}^{2}+\frac{k_{B}}{2}\ln\left(\frac{\kappa_{0}}{\tilde{\kappa}(\tau)}\right)-\frac{\kappa_{0}x_{0}^{2}}{2T}+\frac{\tilde{\kappa}(\tau)x_{\tau}^{2}}{2T},\label{eq:110}
\end{equation}
 and we also have
\begin{equation}
\Delta W=\int_{0}^{\tau}\frac{1}{2}\dot{\kappa}x_{t}^{2}dt.\label{eq:111}
\end{equation}
 We can investigate the statistics of these quantities:
\begin{equation}
\langle\Delta W\rangle=\int_{0}^{\tau}\frac{1}{2}\dot{\kappa}\langle x_{t}^{2}\rangle dt=\int_{0}^{\tau}\frac{1}{2}\dot{\kappa}\frac{k_{B}T}{\tilde{\kappa}}dt,\label{eq:112}
\end{equation}
 and from $\Delta W_{d}=\Delta W-\Delta F$, the rate of performance
of dissipative work is
\begin{equation}
\frac{d\langle\Delta W_{d}\rangle}{dt}=\frac{\dot{\kappa}k_{B}T}{2\tilde{\kappa}}-\frac{dF(\kappa(t))}{dt}=\frac{\dot{\kappa}k_{B}T}{2\tilde{\kappa}}-\frac{\dot{\kappa}k_{B}T}{2\kappa}=\frac{k_{B}T}{2}\dot{\kappa}\left(\frac{1}{\tilde{\kappa}}-\frac{1}{\kappa}\right).\label{eq:113}
\end{equation}
 Whilst the positivity of $\langle\Delta W_{d}\rangle$ is assured
for this process, as a consequence of the Jarzynski equation and the
initial equilibrium condition, the rate of change can be both
positive and negative, according to this result.

\begin{figure}[!htp]
\begin{centering}
\includegraphics[clip,width=100mm]{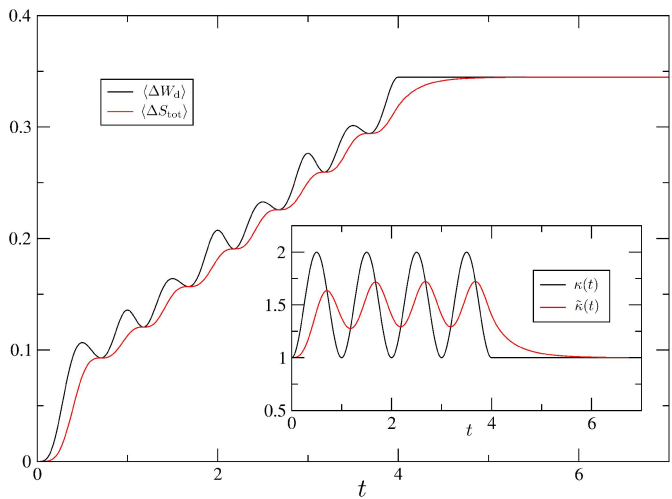} \caption{An illustration of a the mean behaviour of the dissipative work and entropy production for an oscillatory compression and expansion process starting in equilibrium. The mean dissipative work increases, but not monotonically, and is delivered instantly such that there is no change when the protocol stops changing. The mean entropy production however, continues to increase monotonically until it reaches the mean dissipative work after the protocol has stopped changing. The evolution of the protocol, $\kappa(t)=\sin^2(\pi t)+1$, and the characterisation of the distribution, $\tilde{\kappa}(t)$, are shown inset.Units are $k_B=T=m=\gamma=1$.}
\label{mean}
\end{centering}
\end{figure}

\begin{figure}[!htp]
\begin{centering}
\includegraphics[clip,width=100mm]{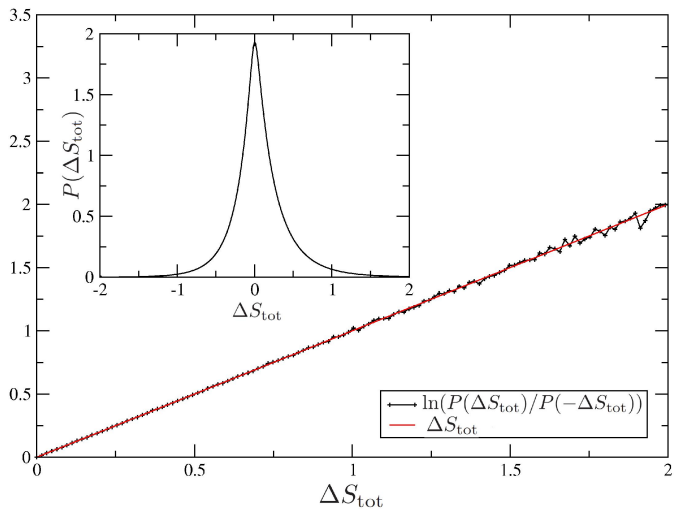} \caption{An illustration of a detailed fluctuation theorem arising for an oscillatory non-equilibrium steady state, as described in section \ref{FTentropy:sec}, created by compressing and expanding a particle in an oscillator by using protocol $\lambda(t)=\kappa(t)=\sin^2(\pi t)+1$. The total entropy production must be measured over a time period during which the protocol is symmetric and the distribution is deemed to be oscillatory. Such a time period exists between $t=3$ and $t=4$ as shown inset in Figure \ref{mean}. Units are $k_B=T=m=\gamma=1$.}
\label{dft}
\end{centering}
\end{figure}

The expectation value for total entropy production in equation (\ref{eq:110}),
on the other hand, is
\begin{equation}
\langle\Delta S_{{\rm tot}}\rangle=\frac{1}{T}\int_{0}^{\tau}\frac{1}{2}\dot{\kappa}\frac{k_{B}T}{\tilde{\kappa}}dt-\frac{1}{2T}\kappa(\tau)\frac{k_{B}T}{\tilde{\kappa}}+\frac{1}{2T}k_{B}T+\frac{k_{B}}{2}\ln\left(\frac{\kappa_{0}}{\tilde{\kappa}(\tau)}\right)-\frac{k_{B}T}{2T}+\frac{k_{B}T}{2T},\label{eq:114}
\end{equation}
 and the rate of change of this quantity is
\begin{align}
\frac{d\langle\Delta S_{{\rm tot}}\rangle}{dt} & =\frac{\dot{\kappa}k_{B}}{2\tilde{\kappa}}-\frac{k_{B}}{2}\left(\frac{\dot{\kappa}}{\tilde{\kappa}}-\frac{\kappa}{\tilde{\kappa}^{2}}\dot{\tilde{\kappa}}\right)-\frac{k_{B}}{2}\frac{\dot{\tilde{\kappa}}}{\tilde{\kappa}}\nonumber \\
 & =\frac{k_{B}}{2}\dot{\tilde{\kappa}}\frac{\left(\kappa-\tilde{\kappa}\right)}{\tilde{\kappa}^{2}}\nonumber \\
 & =\frac{k_{B}}{m\gamma}\frac{\left(\kappa-\tilde{\kappa}\right)^{2}}{\tilde{\kappa}}.\label{eq:115}
\end{align}
 The monotonic increase in entropy with time is explicit. The mean dissipative work and entropy production for a process of this kind starting in equilibrium are illustrated in Figure \ref{mean} where the protocol changes over a driving period followed by a subsequent period of equilibration. Notice particularly that the mean entropy production never exceeds the mean dissipative work, which is delivered instantaneously, and that both take the same value as $t\to\infty$ giving insight into the operation of the Jarzynski equality as discussed in section \ref{workrelations:sec}.
\\\\
It is of more interest however, to verify that detailed fluctuation relations
hold. Analytic demonstration based upon equation (\ref{eq:110}) and
the probability density for a particular trajectory throughout the
entire period is challenging, but a numerical approach based upon
generating sample trajectories is feasible particularly since the entire distribution can always be characterised with the known quantity $\tilde{\kappa}(t)$. As such we may consider the same protocol, $\kappa(t)=\sin^2(\pi t)+1$, wait until the system has reached a non-equilibrium oscillatory steady state as described in section \ref{FTentropy:sec}, characterised here by an oscillatory $\tilde{\kappa}(t)$ as seen in Figure \ref{mean}, and measure the entropy production over a time period across which $\kappa(t)$ is symmetric. The distribution in total entropy production over such a period and the symmetry it possesses are illustrated in Figure \ref{dft}.

\subsection{A simple non-equilibrium steady state}

Let us construct a very simple non-equilibrium steady state. We
consider an overdamped Brownian motion on a ring driven in one direction
by a non-conservative force. We may for sake of argument assume a
constant potential $\phi(x)=c$ such that the equation of motion is
simply
\begin{equation}
\dot{x}=\frac{f}{m\gamma}+\left(\frac{2k_{B}T}{m\gamma}\right)^{1/2}\xi(t).
\end{equation}
 This is just the Wiener process from equation (\ref{eq:33}) centred on a mean proportional to
the external force multiplied by the time so the
probability density of a given displacement is defined by
\begin{equation}
P[x(\tau)|x(0)]=\sqrt{\frac{m\gamma}{4\pi k_{B}T\tau}}\exp{\left[-\frac{m\gamma\left(\Delta x-\frac{f}{m\gamma}\tau\right)^{2}}{4k_{B}T\tau}\right]},
\end{equation}
where the lack of a superscript on $P$ recognises the constancy of
the protocol, and noting that $\Delta x=x(\tau)-x(0)$ may extend
an arbitrary number of times around the ring. Additionally by utilising
the symmetry of the system we can trivially state that the stationary
distribution is given by
\begin{equation}
P^{st}(x)=L^{-1}
\end{equation}
 where $L$ is the circumference of the ring. Let us consider the
nature of the dynamics in this steady state. Considering that we are
in the steady state we know that the transitions must balance in total;
however let us consider the transitions between individual configurations:
comparing the probabilities of transitions we
immediately see that
\begin{equation}
P^{st}(x(0))P[x(\tau)|x(0)]\neq P^{st}(x(\tau))P[x(0)|x(\tau)].
\end{equation}
 Detailed balance explicitly does not hold. For this system not only
can there be entropy production due to driving such as is the case
with expansion and compression processes, but there is a continuous probability
current in the steady state, in the direction of the force, which dissipates heat into the thermal
bath. We have previously  stated in section \ref{FTentropy:sec} that the distribution of the
entropy production in such steady states obeys a detailed fluctuation theorem
for all times. Let's verify that this is the case. The entropy production
is rather simple and is given by
\begin{align}
\Delta S_{{\rm tot}} & =k_{B}\ln{\frac{P^{st}(x(0))P[x(\tau)|x(0)]}{P^{st}(x(\tau))P[x(0)|x(\tau)]}}=k_{B}\ln{\frac{L\exp{\left[-\frac{m\gamma\left(\Delta x-\frac{f}{m\gamma}\tau\right)^{2}}{4k_{B}T\tau}\right]}}{L\exp{\left[-\frac{m\gamma\left(-\Delta x-\frac{f}{m\gamma}\tau\right)^{2}}{4k_{B}T\tau}\right]}}}\nonumber \\
 & =\frac{f\Delta x}{T}.
\end{align}
 This provides an example where the entropy production is highly intuitive.
Taking $f>0$, if the particle moves with the probability current, $\Delta x>0$, it is doing the expected behaviour
and thus is following an entropy generating trajectory. If however,
the particle moves against the current, $\Delta x<0$, it is behaving unexpectedly
and as such is performing a trajectory that destroys entropy. It follows
that since an observation of the particle flowing with a current is more
likely than an observation of the opposite, then on average the
entropy production is positive.\\
\\
 Since the system is in a steady state
we expect a detailed fluctuation theorem. The transformation of the
probability distribution is trivial and we have simply
\begin{equation}
P(\Delta S_{{\rm tot}})=\sqrt{\frac{m\gamma T}{4\pi k_{B}f^{2}\tau}}\exp{\left[-\frac{m\gamma T\left(\Delta S_{{\rm tot}}-\frac{f^{2}}{m\gamma T}\tau\right)^{2}}{4k_{B}f^{2}\tau}\right]}
\end{equation}
 and we can verify a detailed fluctuation theorem which holds for
all time. We can probe further though. Whilst we might conceive of
some fluctuations against a steady flow for a small particle we would
be quite surprised to see such deviations as we approached a macroscopically
sized object. Despite the model's limitations let us consider an approach
to macroscopic behaviour whilst maintaining constant the ratio
$f/m$ such that the mean particle velocity is unchanged. Both the
mean and variance of the distribution of entropy production increase
in proportion. On the scale of the mean, the distribution of entropy
change increasingly looks like a narrower and narrower gaussian until
it inevitably, for a macroscopic object, approaches a delta function
where we recover the classical thermodynamic limit.

\section{Final remarks}

The aim of this chapter was to explore the origin, application and
limitations of fluctuation relations. We have done this within a framework
of stochastic dynamics with white noise and often employing the overdamped
limit in example cases where the derivations are easier: it is in the
analysis of explicit examples where understanding is often to be found.
Nevertheless, the results can be extended to other more complicated stochastic systems, though
the details will need to be sought elsewhere. The fluctuation relations can also be derived within a framework of deterministic, reversible dynamics, which we have discussed briefly in section \ref{determ}. It is interesting to note that within that framework, irreversibility finds its origins in non-linear terms which provide a contraction of phase space, in contrast to the more direct irreversibility found in stochastic descriptions. Both approaches, however, are attempts to represent a dissipative environment that imposes a thermal constraint.

The fluctuation relations concern the statistics of quantities associated
with thermodynamic processes, in particular the mechanical work done
upon, or the heat transferred to a system in contact with a heat bath.
In the thermodynamic limit, the statistics are simple: there are negligible
deviations from the mean, and work and heat transfers appear to be
deterministic and the second law requires entropy change to be non-negative.
But for finite size systems, there are fluctuations, and the statistics
of these will satisfy one or more fluctuation relation. These can
be very specific requirements, for example relating the probability
of a fluctuation with positive dissipative work to the probability
of a fluctuation with negative dissipative work in the reversed process.
Or the outcome can take the form of an inequality that demonstrates
that the mean dissipative work over all possible realisations of the
process is positive.

The core concept in the analysis, within the framework of stochastic dynamics at least, is entropy production. This no longer
need be a mysterious concept: it is a natural measure of the departure
from dynamical reversibility, the loosening of the hold of Loschmidt's
reversibility expectation, when system interactions with a coarse-grained
environment are taken into account. Entropy production emerges in
stochastic models where there is uncertainty in initial specification. Intuitively,
uncertainty in configuration in such a situation will grow with time,
and mean entropy production is this concept commodified. And it turns
out that entropy production can also be related, in certain circumstances, to heat
and work transfers, allowing the growth of uncertainty to be monitored
in terms of thermodynamic process variables. Moreover, although it
is \emph{expected} to be positive, entropy change can be negative,
and the probability of such an excursion, possibly observed by a measurement
of work done or heat transferred, might be described by a fluctuation
relation. In the thermodynamic limit the entropy production appears
to behave deterministically and to violate time reversal symmetry, and
only then does the second law acquire its unbreakable status. But
for small systems, this status is very much diminished, and the second law is revealed
as a statement only about what is likely, within a framework of
rules governing the evolution of probability that explicitly break
time reversal symmetry.

\bibliographystyle{wivchnum}
\bibliography{entropy}
\backmatter

\end{document}